\pdfoutput=1

\documentclass[11pt]{article}
 
\usepackage[]{acl}

\usepackage{times}
\usepackage{latexsym}
\usepackage{xspace}
\usepackage[T1]{fontenc}

\usepackage[utf8]{inputenc}
\usepackage{microtype}
\usepackage[export]{adjustbox}
\usepackage{inconsolata}
\usepackage{amsmath}
\usepackage{booktabs}
\usepackage{multirow}
\usepackage{amsmath}
\usepackage{enumitem}
\usepackage{float}
\usepackage{graphicx}
\usepackage{subcaption}
\usepackage[breakable]{tcolorbox}
\usepackage{colortbl}

\newcommand{\ours}{\textsl{CodeEvo}\xspace}
\newcommand{\ourdata}{\textsl{CodeEvo-100K}\xspace}

\usepackage{tabularray}
\UseTblrLibrary{booktabs}
\SetTblrInner[booktabs]{abovesep=0pt, belowsep=0pt, rowsep=0.2pt}
\SetTblrInner[booktabs]{cells = {cmd=\small}}
\NewTableCommand\seprule{\specialrule{\lightrulewidth,gray8}{2pt}{2pt}}
\NewTableCommand\uniquerule{\specialrule{\lightrulewidth,gray7,dashed}{2pt}{2pt}}
\newcommand{\seprule}{\arrayrulecolor{gray!50}\midrule\arrayrulecolor{black}}
\definecolor{lightb}{RGB}{235,245,255}

\definecolor{gray}{rgb}{0.5,0.5,0.5}

\definecolor{veronica-red}{RGB}{196,30,58}
\definecolor{ForestGreen}{RGB}{34,139,34}
\definecolor{BrickRed}{rgb}{.72,0,0}
\definecolor{LakeBlue}{RGB}{0,61,153}
\usepackage{algorithm}%

\usepackage{algpseudocode}%

\algrenewcommand{\Return}{\textbf{return}~}

\definecolor{lightblue}{RGB}{68,14,196}

\definecolor{arrowblue}{RGB}{68,114,196}
\definecolor{arrowred}{RGB}{192,0,0}
\usepackage{marvosym}

\usepackage{fdsymbol}
\usepackage{fontawesome}
\newcommand{\fstar}{\textsuperscript{\fontsize{6pt}{6pt}\selectfont \faStarO}}
\newcommand{\fmoon}{\textsuperscript{\fontsize{6pt}{6pt}\selectfont \faMoonO}}
\newcommand{\flemon}{\textsuperscript{\fontsize{6pt}{6pt}\selectfont \faLemonO}}

\newcommand{\qwen}{Qwen2.5-32B-Instruct\xspace}

\newcommand{\qwenc}{Qwen2.5-Coder-32B-Instruct\xspace}
\newcommand{\qwencs}{Qwen2.5-Coder-7B-Instruct\xspace}
\newcommand{\qwenthree}{Qwen3-8B\xspace}
\newcommand{\gptoss}{gpt-oss-120B\xspace}

\newcommand{\dpskc}{DeepSeek-Coder-6.7B-Instruct\xspace}
\newcommand{\ivl}{InternLM3-8B-Instruct\xspace}
\newcommand{\starc}{StarCoder2-7B\xspace}

\title{%
  \raisebox{0.075em}{\includegraphics[scale=0.035,valign=c]{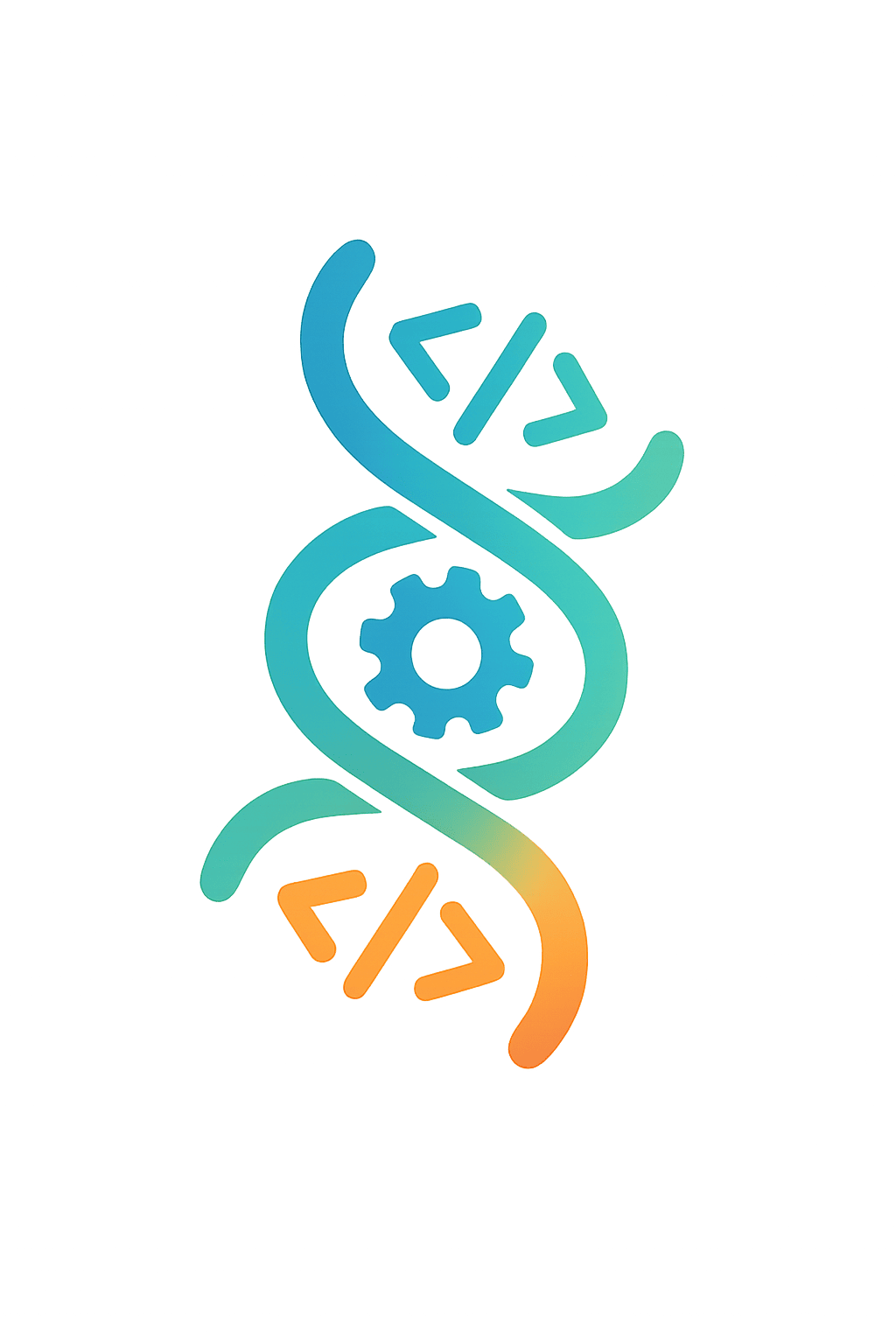}}%
  \hspace{-0.65em}%
  \vspace{-0.95em}
  \ours: Interaction-Driven Synthesis of Code-centric Data through Hybrid and Iterative Feedback%
}

\author{
Qiushi Sun\textsuperscript{$\heartsuit\diamondsuit$}\;
Jinyang Gong\textsuperscript{$\diamondsuit$}\fmoon\;
Lei Li\fstar \quad
\textbf{Qipeng Guo}\textsuperscript{$\diamondsuit$}\flemon \textsuperscript{\Letter}  \;
\textbf{Fei Yuan}\textsuperscript{$\diamondsuit$} \\
\textsuperscript{$\heartsuit$}The University of Hong Kong 
\textsuperscript{$\diamondsuit$}Shanghai AI Laboratory \\
\fmoon New York University 
\fstar Carnegie Mellon University 
\flemon Shanghai Innovation Institute \\
\texttt{qiushisun@connect.hku.hk} \; \texttt{jingyang.gong@nyu.edu}\; \texttt{leili@cs.cmu.edu} \\
\texttt{\{guoqipeng,yuanfei\}@pjlab.org.cn}
}

\begin{document}

\maketitle
\begingroup
\renewcommand\thefootnote{}
\footnotetext{\textsuperscript{\Letter} Corresponding author.}
\endgroup

\begin{abstract}

Acquiring high-quality instruction-code pairs is essential for training Large Language Models for code generation. 
While automated synthesis has emerged as an alternative to expensive manual curation, current approaches often rely on rigid heuristics, yielding data that is ungrounded or lacks logical complexity.
We propose \ours, a dual-agent architecture comprising a Coder for iterative solution synthesis and a Reviewer to orchestrate the generation trajectory. To transcend the limitations of existing heuristics,
the Reviewer formulates a Schema to systematically architect logic and complexity through an interleaved synthesis of instructions and code. This process is further reinforced by a hybrid verification protocol synergizing deterministic compiler feedback with semantic evaluation.
Under this framework, 
we construct \ourdata, a large-scale dataset of instruction–code pairs with stepped difficulty levels.
Extensive experiments demonstrate that models fine-tuned on \ours data consistently outperform established baselines across code generation benchmarks. In-depth analyses further provide insights into effective code-centric data synthesis.
Code and data are available at \url{https://github.com/QiushiSun/CodeEvo}.
\end{abstract}

\section{Introduction}

\begin{figure}[htb]
  \centering
  \includegraphics[width=1\linewidth]{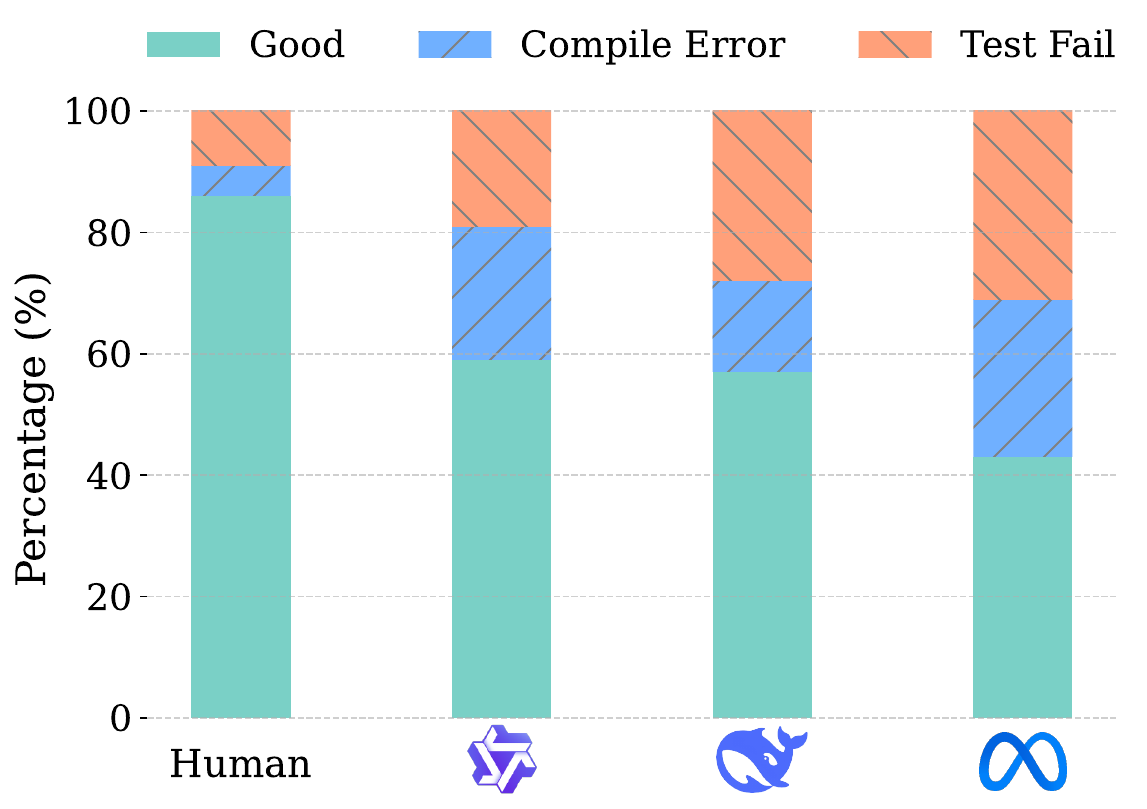}
    \vspace{-2em}
  \caption{Comparison of synthesized code data quality across human examples and code produced by Qwen2.5-Coder-32B, DeepSeek-V3, and Llama-3.1-8B.}
  \label{fig:synthesis_issues}
  \vspace{-1em}
\end{figure}

The rapid development of Large Language Models~(LLMs) has significantly advanced code intelligence~\citep{sun2024survey},
powering applications ranging from line-level code completion to competition-level problem solving.
To further enhance their performance on code generation,
it is essential to train these models with complex, diverse, and grounded instruction-code pairs~\citep{zan2023large}. 
While manually curated data serve as ideal resources, 
their collection is labor-intensive, difficult to scale, and gradually exhausted~\citep{huang2024OpenCoder}.
These limitations have stimulated growing interest in constructing code-centric synthetic data with minimal human intervention.

After early attempts that leverage symbolic augmentation over existing code references~\citep{wang2017program,feng2018program}, 
recent research has shifted toward using LLMs to automatically generate instruction-code pairs. 
These methods, 
such as Self-Instruct~\citep{wang2023selfinstruct} and Evol-Instruct~\citep{luo2024wizardcoder}, 
aim to bootstrap massive data using powerful models and predefined heuristics.
While these approaches enable data construction,
they often fall short in ensuring semantic correctness and executability~\citep{liu2023evalplus}.

As shown in Figure~\ref{fig:synthesis_issues}, we sample instruction-code pairs synthesized by various (Code)LLMs using Evol-Instruct heuristics~\citep{xu2024wizardlm}.
Many of these samples fail to execute or do not pass the provided unit tests, indicating substantial quality gaps. 
These shortcomings can be attributed to two main factors:
(1) instructions are often poorly grounded,
leading to vague or inconsistent objectives; 
and 
(2) generated codes lack proper validation, due to the absence of robust mechanisms to enforce correctness during synthesis.
This motivates a key question:
\textit{Can we design a fully automated and reference-free synthesis pipeline that produces well-grounded and executable instruction-code pairs?
}

Recently emerging LLM agents have demonstrated strong interactive capabilities~\citep{sun2024corex},
enabling them to perform tasks through multi-turn interactions and feedback-driven decision making~(\textit{e.g.}, collaborative programming; \citealp{qian2024chatdev,wang2025openhands}).
These make them promising candidates for moving beyond vanilla data generation toward verifiable and adaptive synthesis pipelines.
Inspired by this potential, 
we propose \ours, an interaction-driven synthesis framework that orchestrates LLM agents to generate high-quality code-centric data. 
Specifically, 
a \textit{Coder} agent produces candidate code and tests based on given instructions, 
while a \textit{Reviewer} agent provides tailored feedback and dynamically constructs new instructions iteratively. 

To address the two core challenges in instruction-code synthesis, 
\ours incorporates two key mechanisms:
(1) To lift instruction quality, we introduce a schema-driven synthesis process. Guided by task-specific keywords, a Reviewer agent conceptualizes a Schema to systematically architect logic and complexity through an interleaved synthesis of instructions and code. This ensures task requirements are intrinsically grounded in functional implementations, transforming instruction evolution from a heuristic-based task into a principled design process.
(2) To boost functional correctness, we introduce a hybrid feedback loop that iteratively refines solutions by fusing the deterministic verification of a compiler with fine-grained semantic judgment of an LLM agent.
The entire pipeline operates with only a small set of seed instructions as input and requires \textit{no human annotation} or \textit{gold references}. Leveraging this framework, 
we construct \ourdata, a large-scale instruction-code dataset featuring stepped difficulty levels and complete evolutionary trajectories.

Experiments across multiple backbones and benchmarks demonstrate that \ours consistently outperforms established data synthesis methods.
Notably,
\ours achieves better performance than competing approaches using several times more data,
indicating the superiority of our targeted, feedback-driven synthesis over sheer data volume.
Our primary contributions are as follows:
\begin{itemize}[itemsep=2pt,topsep=3pt,parsep=0pt]
\item We propose \ours, an interaction-driven synthesis framework that systematically lifts the quality of code-centric synthetic data from both the instruction and code perspectives.
\item A novel hybrid feedback mechanism is designed to synergistically integrate the rigorous determinism of compiler verification with the adaptive generative flexibility of LLM agents.
\item We release a large-scale dataset, \ourdata, featuring {stepped difficulty} levels and capturing the complete {evolutionary trajectory} of the synthesis process.
\item Through extensive experiments and analysis, we share insights into the key attributes, including quality, diversity, scalability, and difficulty of synthetic code data.

\end{itemize}

\begin{figure*}[ht]
  \centering
  \includegraphics[width=\linewidth]{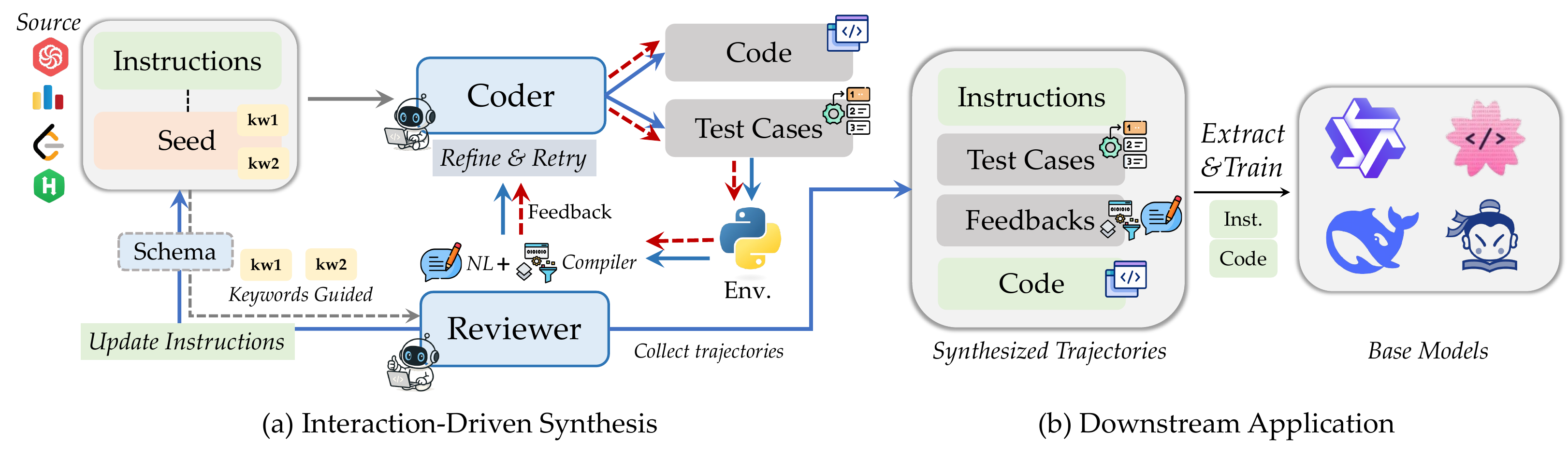}
    \vspace{-1.5em}
\caption{Overview of the \ours framework.
The synthesis process begins with seed instructions from different sources.
Through continuous interaction, the Coder and Reviewer agents collaboratively construct trajectories of instruction, solution, validation, and refinement.
The loop marked with \textcolor{arrowblue}{$\rightarrow$} illustrates the flow of data synthesis, 
where new instructions are derived and paired with validated code.
The loop marked with \textcolor{arrowred}{$\rightarrow$} captures the validation cycle, incorporating natural language and execution feedback to guide refinement.
Instruction-code pairs are extracted from the validated trajectories and used for downstream model training.}
  \label{fig:architecture}
  \vspace{-0.75em}
\end{figure*}

\section{Related Works}

\paragraph{LLM-based Agents Interaction.}

The interactive capabilities of language agents~\citep{sumers2024cognitive},
whether with other agents or the environment, 
have garnered significant attention~\citep{park2023generativeagents}. 
These interactions allow for complex problem-solving approaches such as collaboration~\citep{sun2024corex,hong2024metagpt} or role-playing~\citep{li2023camel,islam2024mapcoder}, which have proven effective in diverse applications like software engineering~\citep{qian2024chatdev,huang2024agentcoder}.
Recently, researchers have begun to explore using such interaction for various data generation,
including instructions~\citep{mitra2024agentins},
reasoning chains~\citep{yang2024react},
and environment-aware trajectories~\citep{qiu2024agent,hu2025agentgen}. 
Leveraging interaction for flexible and scalable data construction is emerging as a promising direction~\citep{luo2024arenalearning,khan2025dataenvgym}. 
This work takes an initial step toward applying interaction to instruction-code synthesis.

\vspace{-0.25em}
\paragraph{Code-centric Data Synthesis.} 
The synthesis of instruction-code data traces back to early symbolic augmentation methods~\citep{wang2017program,feng2018program}, which augment existing code using program transformations.
Recent efforts move beyond static heuristics and leverage LLMs to generate instruction-code pairs at scale with prompting~\citep{codealpaca,wang2023selfinstruct,xu2025magpie}, 
as well as through automated interactions~\citep{wei2024selfcodealign}.
Typically,
WizardCoder~\citep{luo2024wizardcoder} extends Evol-Instruct~\citep{xu2024wizardlm} into code data synthesis,
Magicoder~\citep{wei2024magicoder} and WaveCoder~\citep{yu2024wavecoder} derive pairs from open-source code snippets.
Further, researchers lay emphasis on executable synthesis~\citep{zheng2024opencodeinterpreter,majumdar2025genetic,zeng2025acecoder,xu2025genius}, using code syntax relationships~\citep{wang2025epicoder} and unit tests~\citep{shao2025case2code,ma2025unitcoder} to curate code-centric data.
However, existing synthesis methods often rely on pre-defined and limited prompting heuristics, require existing code references, and largely overlook functional correctness.
\vspace{-.75em}

\paragraph{Compiler Feedback in Code Generation.}
A key differentiator between symbolic language and natural language is executability~\citep{xu2024symbol}, 
with compilers serving as a fundamental verifier~\citep{wang2022compcoder}. In the context of LLM-based code generation, leveraging compiler feedback to improve output quality has become an active area of research.
Early approaches primarily focused on post-hoc error correction using compiler signals~\citep{lahiri2022ticoder}, later evolving to incorporate immediate compiler feedback during generation to improve first-pass correctness~\citep{wang2022compcoder}.
Static analysis has also been explored to enrich the semantic understanding of code~\citep{ma2024lmsunderstandingcodesyntax}.
More recent efforts have begun decomposing complex generation tasks into subtasks, using compiler feedback for fine-grained optimization~\citep{xu2024envisions},
debugging~\citep{islam2025codesim},
or repo-level learning~\citep{bi2024cocogen}. 
In addition, compiler feedback has also been leveraged to model preferences~\citep{zhang2024codedpo}.
While compilers are widely used in the generation stage, their role in the data synthesis pipeline is underexplored. In our synthesis loop, an LLM agent interprets execution feedback to ensure data quality and guide refinement.
\section{\ours}

We detail the framework of \ours in this section,
emphasizing its core components and systematic workflow, as illustrated in Figure~\ref{fig:architecture}.

\subsection{Preliminary}
\paragraph{Problem Definition.} Given a seed dataset $S$,
containing initial instructions $s$,
each with an associated set of keywords $T$.
The goal is to synthesize an expanded dataset $Q$.
\ours adopts a dual-agent architecture with collaborative roles:
\begin{itemize}[itemsep=2.5pt,topsep=3pt,parsep=0pt]
\item \textbf{Coder}: Generate candidate solutions and test cases for a given instruction, and refine its solution based on external feedback.
\item \textbf{Reviewer}: Generate new instructions and evaluate candidate solutions, providing feedback for refinement or data selection.
\end{itemize}

\paragraph{Seed Instructions.}
In contrast to prior work that relies on golden code solutions or ready-made test cases,
\ours requires \textit{only a lightweight set of natural language instructions}, 
which can originate from any domain where the problem descriptions admit symbolic solutions,
such as algorithmic problems from programming platforms, NL2Code training sets, or mathematically structured problems.

\subsection{Schema-Guided Instruction Generation}
\label{sec:schema}

A central challenge in instruction synthesis is maintaining semantic control during evolution. Prior methods often rely on abstract commands (\textit{e.g.}, ``make it harder''), which can lead to vague or ungrounded content. To move beyond such abstract heuristics, we introduce Schema-guided interleaved synthesis.
This approach transforms synthesis into a principled design process where instructions and code co-evolve.
Guided by task-specific keywords, our Reviewer agent first generates a Schema—a structured blueprint that outlines the logic for combining concepts, the desired complexity, and the overall goal for a new instruction. By first formulating this design plan before generating the final text, we ensure that new instructions are not only more challenging but also logically coherent and well-grounded.

\begin{figure}[ht]
  \centering
  \includegraphics[width=0.9\linewidth]{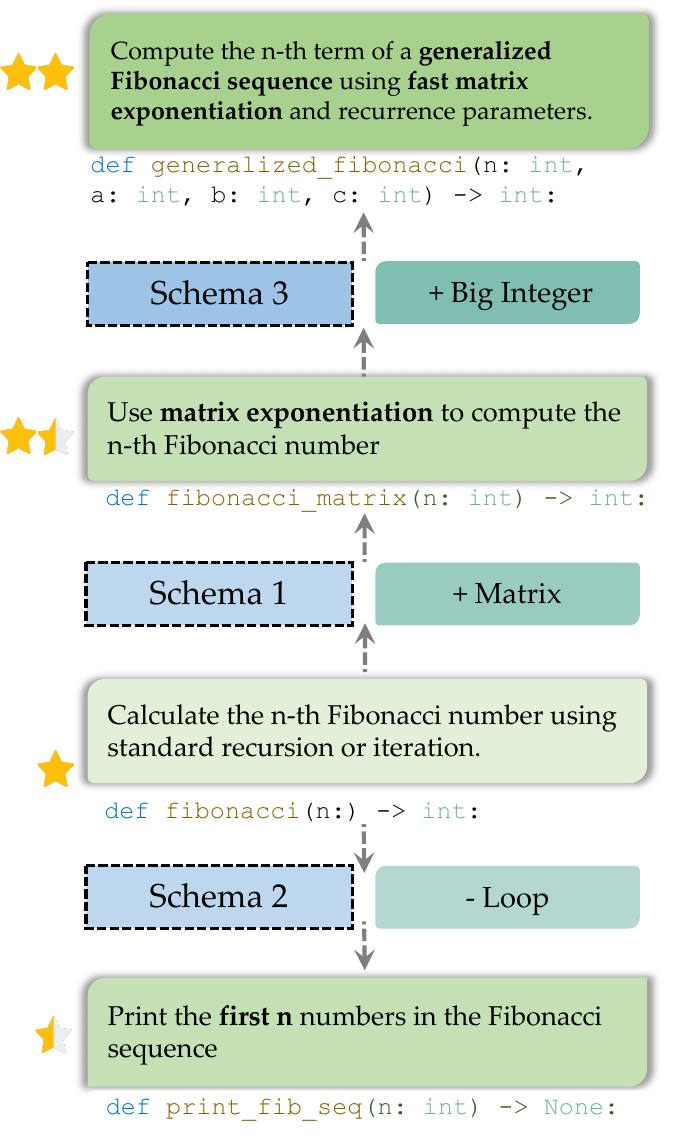}
  \vspace{-0.25em}
  \caption{Illustration of transforming a seed into relevant instructions by leveraging schema and keywords.}
  \label{fig:kw_guide}
\end{figure}

Specifically, for the current instruction $q$ (initialized to a seed instruction $s$) and a selected keyword subset $t \subseteq T$, the generation process unfolds in two stages. First, the Reviewer agent formulates a Schema that serves as a detailed plan for the new instruction:
\vspace{-5pt}
\begin{equation}
\label{eq:schema}
\text{Schema} = \text{GenerateSchema}(q, t)
\end{equation}
\vspace{-10pt}

Subsequently, it generates the new instruction $q^{+}$ by executing the plan laid out in the Schema:
\vspace{-5pt}
\begin{equation}
\label{eq:write}
q^{+} = \text{GenInstruction}(q, \text{Schema})
\end{equation}
\vspace{-10pt}

Crucially, this schema-driven mechanism enables stepped difficulty levels through its bidirectional nature. 
The schema can strategically plan to either integrate keywords for added complexity or selectively omit them to construct a simplified variant, 
particularly when a task proves intractable for the Coder agent.
This structured approach marks a significant shift from rigid prompting heuristics: 
enabling the generation of novel, diverse instructions while reducing the likelihood of producing unanswerable ones (statistics in Appendix~\ref{app:ext_analysis}). An adaptive keyword sampling strategy is detailed in Appendix~\ref{app:seed_and_keywords}.

\subsection{Hybrid Feedback for Validation and Refinement}

A key design of \ours is our hybrid feedback mechanism, 
which synergizes deterministic compiler evaluations with agentic semantic assessments to ensure the functional correctness and logical integrity of synthesized solutions. 
Given an instruction $q$, the Coder first generates a candidate solution $c$ along with a suite of initial or self-generated test cases $g$:

\begin{equation}
\left(c, g, f_{\text{comp}}\right) = \operatorname{Coder}\left(q\right)
\end{equation}

While $f_{\text {comp}}$ provides a deterministic pass/fail signal from the execution of $g$, its reliability is inherently constrained by the quality of the self-generated tests. Considering that synthesized test cases may suffer from limited coverage or internal logical errors, relying solely on execution results often leads to false positives.

To mitigate this, our Reviewer agent acts as a contextual judge to perform a deep semantic audit. Beyond the raw execution signal, it produces a nuanced natural language evaluation $f_{\mathrm{NL}}$ that scrutinizes the solution's logical alignment with the instruction, the implementation of complex constraints, and potential edge cases overlooked by the test suite $g$. These dual signals are then fused into a comprehensive hybrid feedback:

\begin{equation}
f_{\text{hybrid}} = \operatorname{Reviewer}\left(f_{\text{comp}}, f_{\mathrm{NL}}\right)
\end{equation}

The hybrid feedback $f_{\text{hybrid}}$ serves a dual role in \ours: (1) Data Selection: determining whether the instruction-code pair meets the stringent quality bar for inclusion in the final dataset; and (2) Iterative Refinement: providing a rich supervisory signal that guides the Coder to rectify subtle bugs or logical misalignments. This collaborative verification ensures that the final entries in the evolutionary trajectory are both executable and semantically grounded, even without relying on external labels or human-crafted references.

\begin{algorithm}[!ht]
\caption{Interaction-Driven Synthesis}
\label{alg:codeevo-seed}
\small
\begin{algorithmic}[1]
\Require Seed instruction $s$, keyword set $T$, maximum iterations $N=4$, refinement budget $R=3$
\Ensure Validated instruction-code pairs $Q_s$
\State Initialize $Q_s \leftarrow \emptyset$, $q \leftarrow s$, $k \leftarrow 0$

\While{$k < N$}
    \State $(c, g, f_{\text{comp}}) \leftarrow \text{Coder}(q)$
    \State $f_{\text{NL}} \leftarrow \text{Reviewer}(q, c, g, f_{\text{comp}})$
    \State $f_{\text{hybrid}} \leftarrow \text{Reviewer}(f_{\text{comp}}, f_{\text{NL}})$
    \State $r \leftarrow 0$
    \While{$f_{\text{hybrid}}$ invalid \textbf{and} $r < R$} \Comment{Refine \& retry}
        \State $(c, g, f_{\text{comp}}) \leftarrow \text{Coder.Refine}(c, f_{\text{hybrid}})$
        \State $f_{\text{NL}} \leftarrow \text{Reviewer}(q, c, g, f_{\text{comp}})$
        \State $f_{\text{hybrid}} \leftarrow \text{Reviewer}(f_{\text{comp}}, f_{\text{NL}})$
        \State $r \leftarrow r + 1$
    \EndWhile
    \If{$f_{\text{hybrid}}$ is valid}
        \State Add $(q, c)$ to $Q_s$
        \State Sample keywords $t \subseteq T$
        \State $\text{Schema} \leftarrow \text{Reviewer.Plan}(q, t)$; \hspace{2mm} $q^+ \leftarrow \text{Reviewer.Write}(q, \text{Schema})$
        \State $q \leftarrow q^+$
        \State $k \leftarrow k + 1$
    \Else
        \State Sample keywords $t' \subseteq T$
        \State $\text{Schema} \leftarrow \text{Reviewer.Plan}(q, t')$; \hspace{2mm} $q^- \leftarrow \text{Reviewer.Write}(q, \text{Schema})$ \Comment{Plan a simpler variant}
        \State $(c, g, f_{\text{comp}}) \leftarrow \text{Coder}(q^-)$
        \State $f_{\text{NL}} \leftarrow \text{Reviewer}(q^-, c, g, f_{\text{comp}})$
        \State $f_{\text{hybrid}} \leftarrow \text{Reviewer}(f_{\text{comp}}, f_{\text{NL}})$
        \If{$f_{\text{hybrid}}$ is valid}
            \State Add $(q^-, c)$ to $Q_s$
        \EndIf
        \State \textbf{break}
    \EndIf
\EndWhile
\State \Return $Q_s$

\end{algorithmic}

\end{algorithm}

\subsection{Interaction-Driven Synthesis}

With these mechanisms, \ours orchestrates a collaborative refinement loop between the Coder and Reviewer. This transforms data synthesis from a static, one-shot pipeline into an adaptive process where tasks are proposed, attempted, and rigorously assessed. Crucially, the framework can dynamically adjust difficulty: if a task proves too challenging, the Reviewer can formulate a simpler Schema to generate a more tractable problem (as shown in Algorithm~\ref{alg:codeevo-seed}).
This self-correcting loop of validation and refinement provides intrinsic quality control, which is key to maintaining a high yield of valid data.
This generates a rich interaction trajectory capturing the full cycle of problem-solving, feedback, and correction, from which high-quality instruction-code pairs can be extracted.
\section{Experiments}
\begin{figure*}[htb]
  \vspace{-1em}
  \centering
  \begin{subfigure}[t]{0.72\textwidth}
    \centering
    \includegraphics[width=\linewidth]{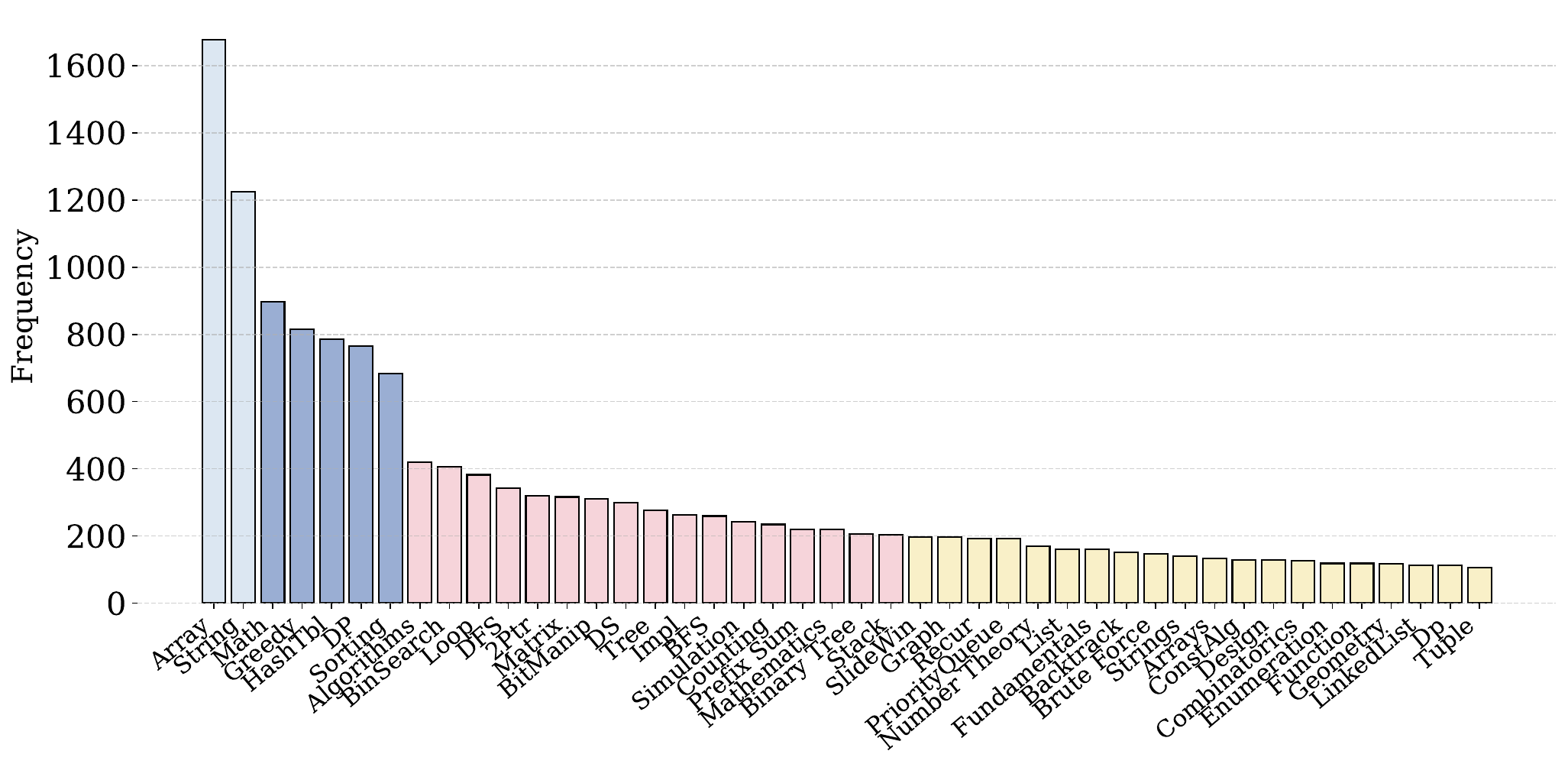}
    \vspace{-1.75em}
    \caption{Distribution of keywords with occurrence frequency $\geq$ 100.}
    \label{fig:keyword-bar}
  \end{subfigure}
  \hfill
  \begin{subfigure}[t]{0.27\textwidth}
    \centering
    \raisebox{7mm}{ 
      \includegraphics[width=\linewidth]{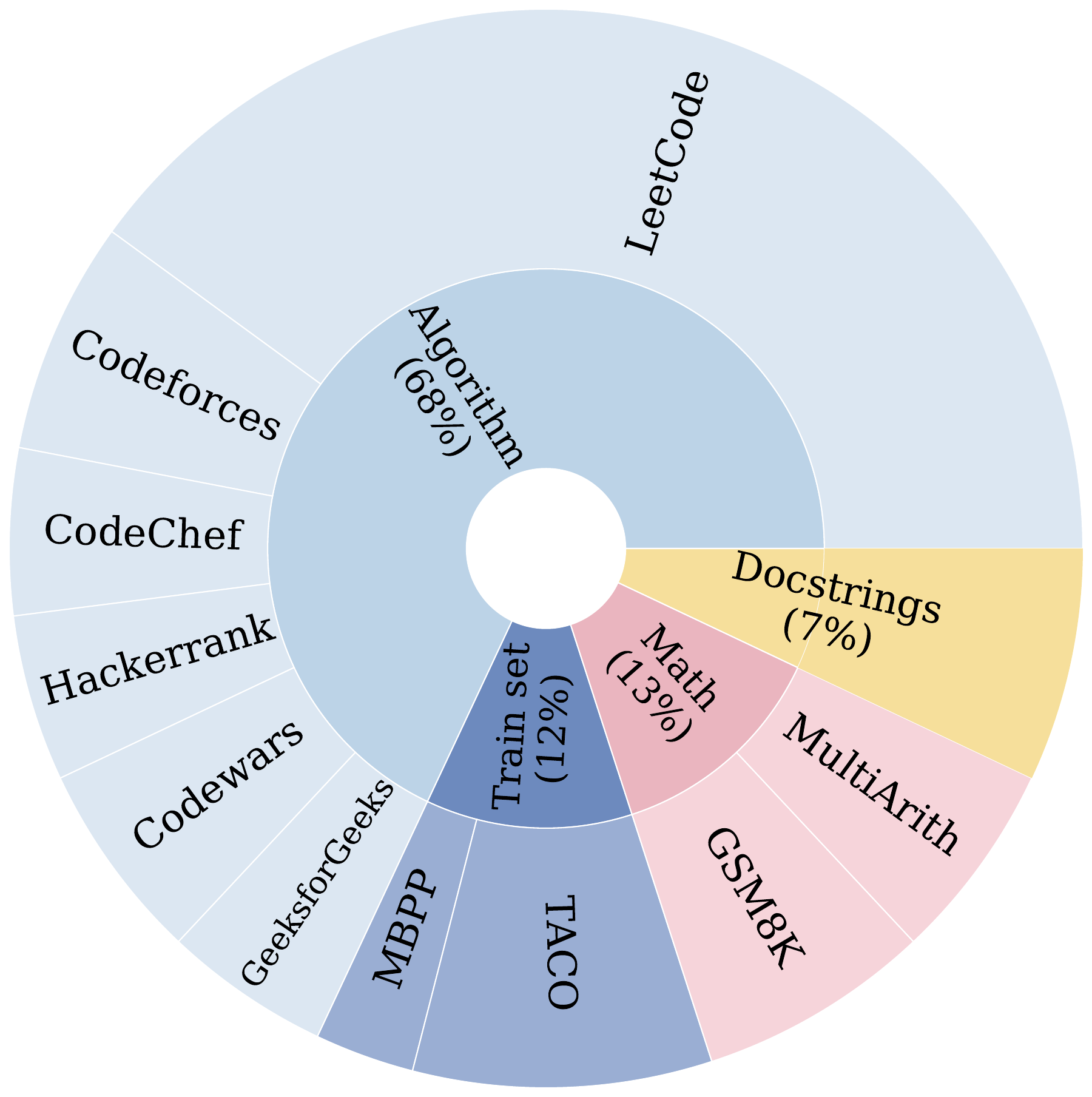}
    }
    \vspace{-1.75em}
    \caption{Type and source of seeds.}
    \label{fig:seed-pie}
  \end{subfigure}
  \caption{An overview of seed instructions and keyword distribution.}
  \label{fig:seed_info}
  \vspace{-1em}
\end{figure*}

\subsection{Experimental Settings}

\paragraph{Model Settings.}
We evaluate our approach under two backbone agent scales: a medium-scale setting with moderately sized coder and reviewer agents, 
and a large-scale setting.
In the medium-scale configuration, we employ \qwenc~\citep{qwen25coder} as the coder agent and \qwen~\citep{qwen25} as the reviewer agent within the data synthesis pipelines of \ours. For the large-scale configuration, we adopt \gptoss~\citep{openai2025gptoss} as both the coder and reviewer agents.
To maximize the synthesis yield,
we utilize the large-scale setting to construct \ourdata dataset. For comparative analysis, we additionally synthesize a dataset subset using the medium-scale setting,
equivalent in scale to the hard set of \ourdata.

\vspace{-0.5em}
\paragraph{Fine-tuning Settings.}
To assess the performance introduced by \ours,
we conduct experiments primarily on \qwenthree~\citep{qwen3}, representing general-purpose LLMs, and \qwencs~\citep{qwen25coder}, representing specialized CodeLLMs. Results on additional backbones are provided in Appendix~\ref{app:ext_results}.
All models are trained with full fine-tuning on clusters of 8 $\times$ A100 80GB GPUs,
with further implementation details provided in Appendix~\ref{app:training_details}. 

\vspace{-0.5em}
\paragraph{Evaluation Benchmarks.}

We evaluate the Python code generation capability of models trained on \ours data using HumanEval~\citep{chen2021codex}, MBPP~\citep{austin2021program}, and their plus versions from EvalPlus~\citep{liu2023evalplus}.
To further assess generalization under realistic difficulty levels, 
we also use 
BigCodeBench~\citep{zhuo2025bigcodebench} and LiveCodeBench~\citep{jain2025livecodebench}
which include more complex instructions, algorithmic logic, and function calls.
Further experimental details can be found in Appendix~\ref{app:exp_details}.

\subsection{Baseline Construction}
\paragraph{Baselines.}
As a pioneering study in synthesizing code-centric data,
we leverage the following baselines to demonstrate the superiority of \ours. 
\begin{itemize}[itemsep=2pt,topsep=3pt,parsep=0pt]
    \item \textbf{Zero-Shot}: The original evaluation setting using zero-shot prompting.
    \item \textbf{Code Evol-Instruct}: Proposed by WizardCoder~\citep{luo2024wizardcoder}, Code Evol-Instruct first evolves task complexity and then generates the code solutions via prompting. 
    As the original dataset is not publicly available, 
    we reproduce this baseline under the same setting as \ours,
    Specifically, we employ \qwen and \qwenc for the medium-scale setting, and \gptoss for the large-scale configuration.
    The same seed data as \ours is used to ensure a fair comparison.\footnote{We refer to this setting as \texttt{Evol-Instruct} in the following experiments for brevity.} 
    \item \textbf{OSS-Instruct}: Released alongside Magicoder~\citep{wei2024magicoder}, OSS-Instruct derives instructions from open-source code snippets written by humans and includes 75K data.
\end{itemize}
Details of the baseline construction are provided in Appendix~\ref{app:baseline}. All these data and resources will be made public to accelerate future research.

\begin{table*}[!h]
\centering
\resizebox{\textwidth}{!}{
\begin{tabular}{lr|cc|cc|cc|cc|c}
\toprule
\multirow{2}{*}{\textbf{Method}} & 
\multirow{2}{*}{\textbf{Data Scale}} & 
\multicolumn{2}{c|}{\textbf{HumanEval}} & 
\multicolumn{2}{c|}{\textbf{MBPP}} &
\multicolumn{2}{c|}{\textbf{BigCodeBench-Full}} &
\multicolumn{2}{c|}{\textbf{BigCodeBench-Hard}} & 
\textbf{LiveCodeBench} \\
& & 
\textit{HE} & \textit{HE+} & 
\textit{MBPP} & \textit{MBPP+} & 
\textit{Instruct} & \textit{Complete} &
\textit{Instruct} & \textit{Complete} &
\textit{v6} \\
\midrule
\raisebox{0.075em}{\includegraphics[scale=0.058,valign=c]{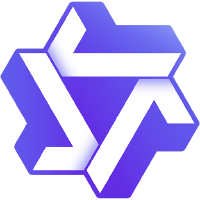}}\textit{\qwencs} & - & 84.1 & \textbf{79.9} & 79.1 & 66.7 & 40.4 & 48.8 & 18.2 & 21.6 & 17.1 \\
\;OSS-Instruct & 75K & 83.5 & 78.0 & 78.0 & 64.8 & 41.4 & 48.6 & \textbf{20.3} & 20.3 & 18.9 \\
\seprule
\multicolumn{11}{c}{Backbones: Qwen2.5-Coder-32B-Instruct | Qwen2.5-32B-Instruct} \\
\seprule
\;Evol-Instruct & 25K & 83.5 & 78.0 & 79.1 & 66.9 & 40.6 & 51.4 & 15.5 & 22.3 & 14.5 \\
\rowcolor{lightb} \;\ours & 17K & 85.3 & 79.9 & 81.2 & 68.5 & 41.9 & 52.2 & 17.6 & 26.4 & 22.3 \\
\seprule
\multicolumn{11}{c}{Backbones: gpt-oss-120B | gpt-oss-120B} \\
\seprule
\;Evol-Instruct & 25K & 85.1 & 80.5 & 81.2 & 69.7 & 42.0 & 50.3 & 19.7 & 23.6 & 22.5 \\
\rowcolor{lightb} \;\ours & 17K & \textbf{86.4} & 80.9 & 83.0 & \textbf{73.2} & \textbf{43.4} & 50.1 & 21.5 & 22.9 & 24.3 \\
\rowcolor{lightb} \;\ours & 100K & 85.7 & \textbf{81.1} & \textbf{84.7} & 72.5 & 42.8 & \textbf{52.1} & \textbf{22.0} & \textbf{24.1} & \textbf{28.1} \\
\midrule
\raisebox{0.075em}{\includegraphics[scale=0.058,valign=c]{figures/icons/qwen.png}}\textit{\qwenthree} & - & 82.9 & 77.4 & 80.7 & 70.9 & 42.7 & 49.2 & 14.9 & 22.3 & 39.1 \\
\;OSS-Instruct & 75K & 84.1 & 78.5 & 79.8 & 67.5 & 42.9 & 50.3 & 15.7 & 24.1 & 36.3 \\
\seprule
\multicolumn{11}{c}{Backbones: Qwen2.5-Coder-32B-Instruct | Qwen2.5-32B-Instruct} \\
\seprule
\;Evol-Instruct & 25K & 79.2 & 74.6 & 77.5 & 67.2 & 41.5 & 47.7 & 12.3 & 20.9 & 36.1 \\
\rowcolor{lightb} \;\ours & 17K & 83.7 & 76.4 & 81.7 & 72.9 & 42.9 & 50.3 & 14.7 & 21.1 & 39.8 \\
\seprule
\multicolumn{11}{c}{Backbones: gpt-oss-120B | gpt-oss-120B} \\
\seprule
\;Evol-Instruct & 25K & 84.5 & 78.2 & 82.4 & 72.5 & 44.1 & 53.3 & 16.5 & 25.2 & 40.9 \\
\rowcolor{lightb} \;\ours & 17K & \textbf{86.7} & 79.8 & 85.5 & \textbf{74.1} & \textbf{44.9} & 52.5 & \textbf{18.3} & 24.1 & 42.7 \\
\rowcolor{lightb} \;\ours & 100K & 86.4 & \textbf{81.7} & \textbf{86.2} & 73.8 & \textbf{44.9} & \textbf{55.1} & 17.6 & \textbf{26.1} & \textbf{46.9} \\

\bottomrule
\end{tabular}
}
\caption{Results of pass@1(\%) performance on various models on HumanEval(+), MBPP(+), BigCodeBench-Full, BigCodeBench-Hard, and LiveCodeBench. All results are from single runs with the fixed seeds described in Appendix~\ref{app:training_details}.}
\label{tab:main_python}
\end{table*}

\subsection{Data Resources}

\paragraph{Seed Instructions.}
We curate a set of $\approx$ 5K seed instructions from diverse sources, including programming platforms such as LeetCode and Codeforces, as well as existing math and code training sets~\citep{cobbe2021gsm8k,austin2021program}. 
Each seed is independently utilized to derive both instruction and completion data formats.
A subset of these instructions is collected alongside their corresponding reference solutions; following \citet{luo2024wizardcoder}, we include them during training with appropriate ablation.

Each seed instruction is paired with a set of keyword tags (averaging 3 per instruction),
which are either inherited from the original sources (\textit{e.g.}, LeetCode tags, MBPP annotations) or assigned automatically when unavailable. 
Figure~\ref{fig:seed_info} provides an overview of the seed data used in our experiments.
Additional statistics, balanced sampling strategy, and implementation details are in Appendix~\ref{app:seed_and_keywords}.

\vspace{-0.5em}
\paragraph{\ourdata.}
Through \ours, 
we construct \ourdata, a large dataset featuring {stepped difficulty levels} based on evolution depth.
We categorize the data into three tiers: (1) \textit{Easy}: data from original instruction and 1-round evolutions; (2) \textit{Medium}: 2-round evolutions; and (3) \textit{Hard}: a high-complexity subset surviving $\ge 3$ evolutionary cycles. Unlike static repositories, \ourdata preserves the complete {evolutionary trajectory}, encompassing intermediate reasoning and refinement iterations.

\subsection{Main Results}

\paragraph{Performance Gains across Benchmarks.}

As shown in Table~\ref{tab:main_python}, models fine-tuned on \ours data outperform the baselines on the majority of benchmarks. This strong performance is evident across data synthesized from both our medium-scale (32B) and large-scale (120B) agent configurations, highlighting the framework's robustness and scalability.

Remarkably, our schema-driven methodology empowers even moderate-scale models to synthesize high-utility data. A subset of merely 17K hard data generated by our 32B agents elicits a larger performance boost for Qwen3-8B than the 75K OSS-Instruct dataset. On LiveCodeBench in particular, OSS-Instruct falls below the zero-shot baseline, whereas \ours preserves it.
Furthermore, expanding to the full \ourdata brings further gains, concentrated on the harder benchmarks (\textit{e.g.}, LiveCodeBench).
This demonstrates that a superior synthesis algorithm can be more critical than the sheer volume of data.

Interestingly, larger gains can be observed on HE+ and MBPP+, which feature extra test cases, suggesting that our ``compiler-in-the-loop'' design in the hybrid feedback plays a critical role in validating functional correctness.

\paragraph{Superior Data Efficiency.}

Despite using fewer training examples, our hard set consistently outperforms other approaches, which rely on 4–5$\times$ more synthetic data and code references.
It further underscores the importance of quality-aware code data construction.

The efficiency stems from innovations on both sides of the pipeline: instruction synthesis is grounded through keyword-driven refinement, while code synthesis is constrained by hybrid feedback. \ours inherently reduces the production of invalid or redundant samples, paving the way for more data-efficient enhancement of code generation capabilities. Further discussions on the impact of data scale are presented in Appendix~\ref{sec:analysis_scaling}.

\paragraph{Ablation Studies.}

\begin{table}[!h]
\centering
\resizebox{\columnwidth}{!}{
\begin{tabular}{lc|c|cc}
\toprule
\multirow{2}{*}{\textbf{Method}} & \multirow{2}{*}{\textbf{HE+}} & \multirow{2}{*}{\textbf{MBPP+}} & \multicolumn{2}{c}{\textbf{BigCodeBench-Full}} \\
& & & \textit{Instruct} & \textit{Complete} \\
\midrule
\raisebox{0.075em}{\includegraphics[scale=0.058,valign=c]{figures/icons/qwen.png}}\textit{\qwencs} \\
\rowcolor{lightb} \;\ours \textit{w/o Seed} & 80.7 & 71.9 & 42.7 & \textbf{51.3} \\
\rowcolor{lightb} \;\ours & \textbf{80.9} & \textbf{73.2} & \textbf{43.4} & 50.1 \\
\midrule
\raisebox{0.075em}{\includegraphics[scale=0.058,valign=c]{figures/icons/qwen.png}}\textit{\qwenthree} \\
\rowcolor{lightb} \;\ours \textit{w/o Seed} & \textbf{80.1} & 73.9 & 44.1 & \textbf{52.9} \\
\rowcolor{lightb} \;\ours & 79.8 & \textbf{74.1} & \textbf{44.9} & 52.5 \\
\bottomrule
\end{tabular}
}
\caption{Ablation study. Using 17K hard data generated through agents backboned by \gptoss. All results are from single runs with the fixed seeds described in Appendix~\ref{app:training_details}.}
\label{tab:seed_ablation_complete}
\end{table}

We perform an ablation study to isolate the effect of seed data.
The results in Table~\ref{tab:seed_ablation_complete} demonstrate that its exclusion does not lead to a significant performance degradation. 
In certain scenarios, models trained exclusively on \ours-synthesized data achieve even superior performance. 
This further suggests that the varied and high-quality data synthesized by \ours provides a more effective training signal than the original seed set, which is inherently more limited in its diversity and scope.

\section{Analysis}
\label{sec:analysis}

Beyond performance gains, we conduct a series of analyses to provide insights into the quality, diversity, scalability, and utility of synthetic code data.

\subsection{Synthetic Data Diversity}
\label{sec:diversity_analysis}

A core feature of \ours is the generation of diverse instructions and preventing overfitting to narrow problem types. 
To assess this, we perform a comparative diversity analysis of instruction samples and instruction-code pairs (N=1000).
We compute the average pairwise cosine similarity over code embeddings \footnote{We leverage \texttt{text-embedding-3-small} to obtain embeddings.} to assess diversity.

\begin{figure}[htb]
  \centering
  \includegraphics[width=\linewidth]{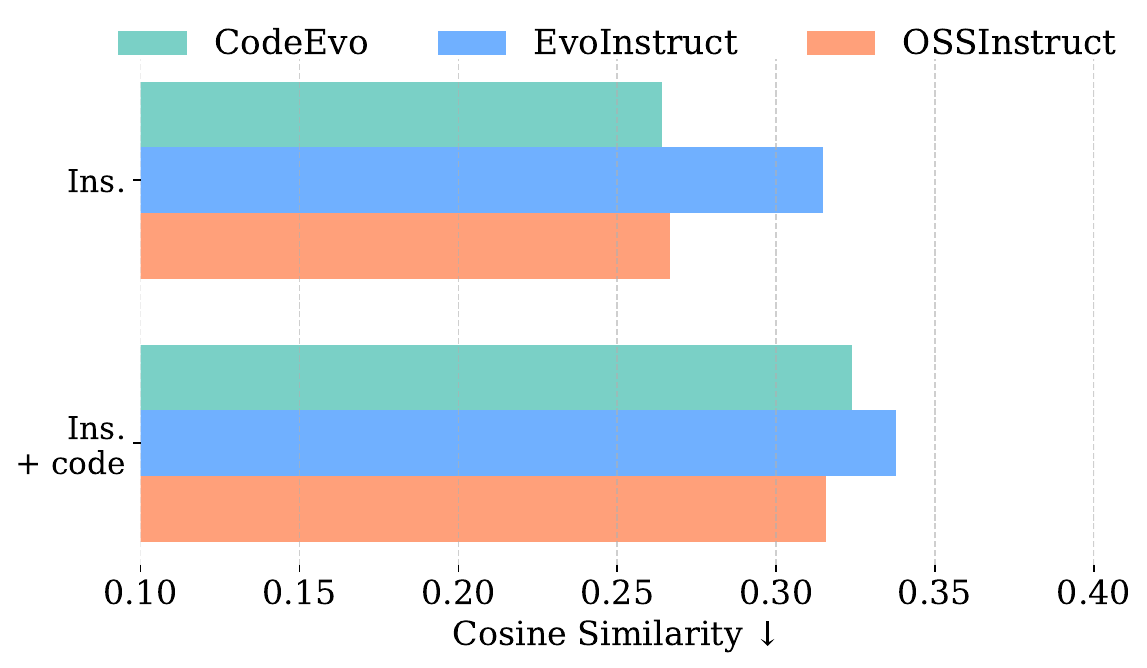}
    \vspace{-0.5em}
  \caption{Comparison of instruction diversity and instruction-code pairs diversity among different synthetic methods.}
  \label{fig:diversity}
  \vspace{-1em}
\end{figure}

As shown in Figure~\ref{fig:diversity}, \ours achieves the lowest average similarity among instruction samples,
demonstrating the effectiveness of our keyword-guided strategy in constructing semantically diverse prompts.
For instruction-code pairs, the diversity of \ours is also comparable to OSS-Instruct, 
which derives data directly from human-written code.
This indicates that, despite undergoing a rigid filtering process, our synthesized data retains a high level of overall diversity.

\subsection{Instruction Difficulty}
\label{sec:harder_ins}
To evaluate whether \ours really generates more challenging instructions, 
we conducted a human study comparing three variants derived from the same seed: 
the original seed instruction, 
an ``evolved'' version from Evol-Instruct, 
and the instruction synthesized by \ours.
Five participants with programming experience are invited to rate the perceived difficulty of each instruction on a scale from 1 (very easy) to 5 (very difficult).

\begin{figure}[htb]
  \centering
  \includegraphics[width=1\linewidth]{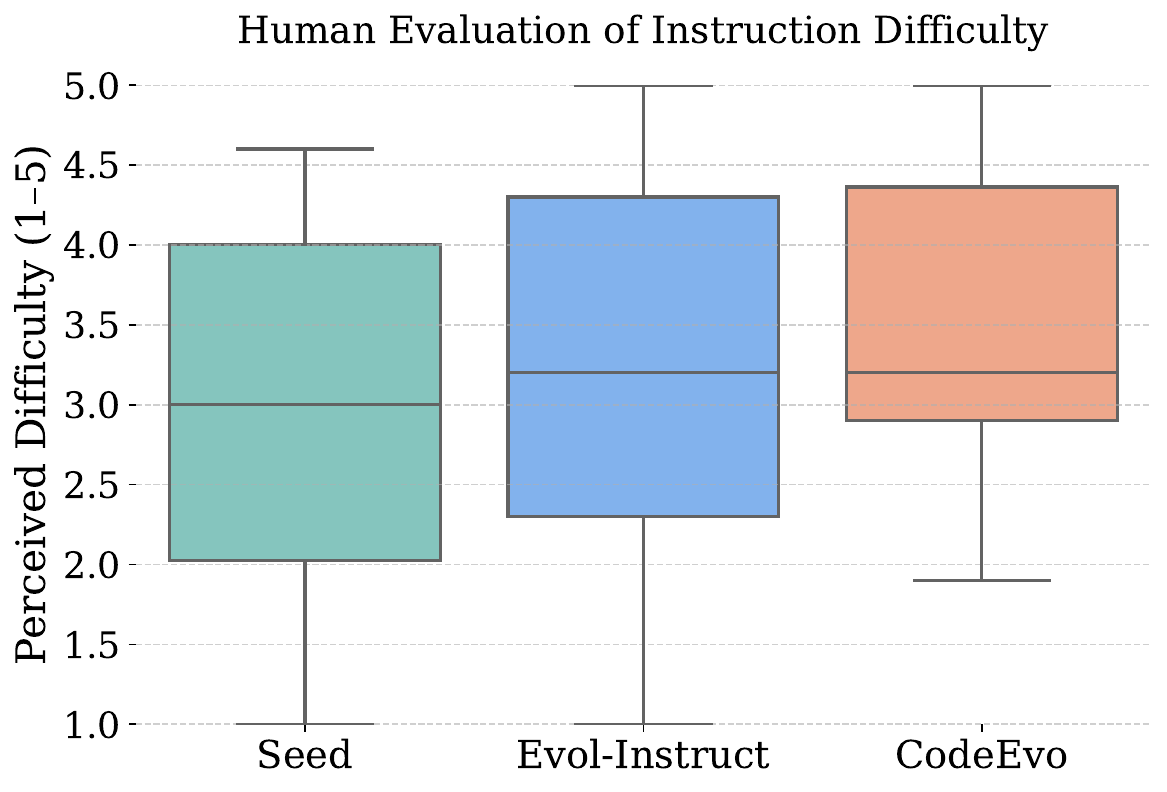}
    \vspace{-1.5em}
  \caption{Human-rated difficulty of instructions. }
  \label{fig:human_difficulty}
  \vspace{-0.5em}
\end{figure}

As shown in Figure~\ref{fig:human_difficulty}, 
\ours instructions (from 32B settings) received the highest difficulty scores (mean $\approx$ 3.5), 
followed by Evol-Instruct and Seed. 
In addition, \ours exhibits a higher lower bound, indicating that the generated instructions are not only more difficult on average, but also more consistently fall within a higher difficulty range,
which further validates the edge of using keyword guidance as a signal.
In contrast, Evol-Instruct fails to consistently increase instruction difficulty.

\subsection{Data Survival Rate}
\label{sec:survival_rate}

\begin{figure}[htb]
  \centering
  \includegraphics[width=\linewidth]{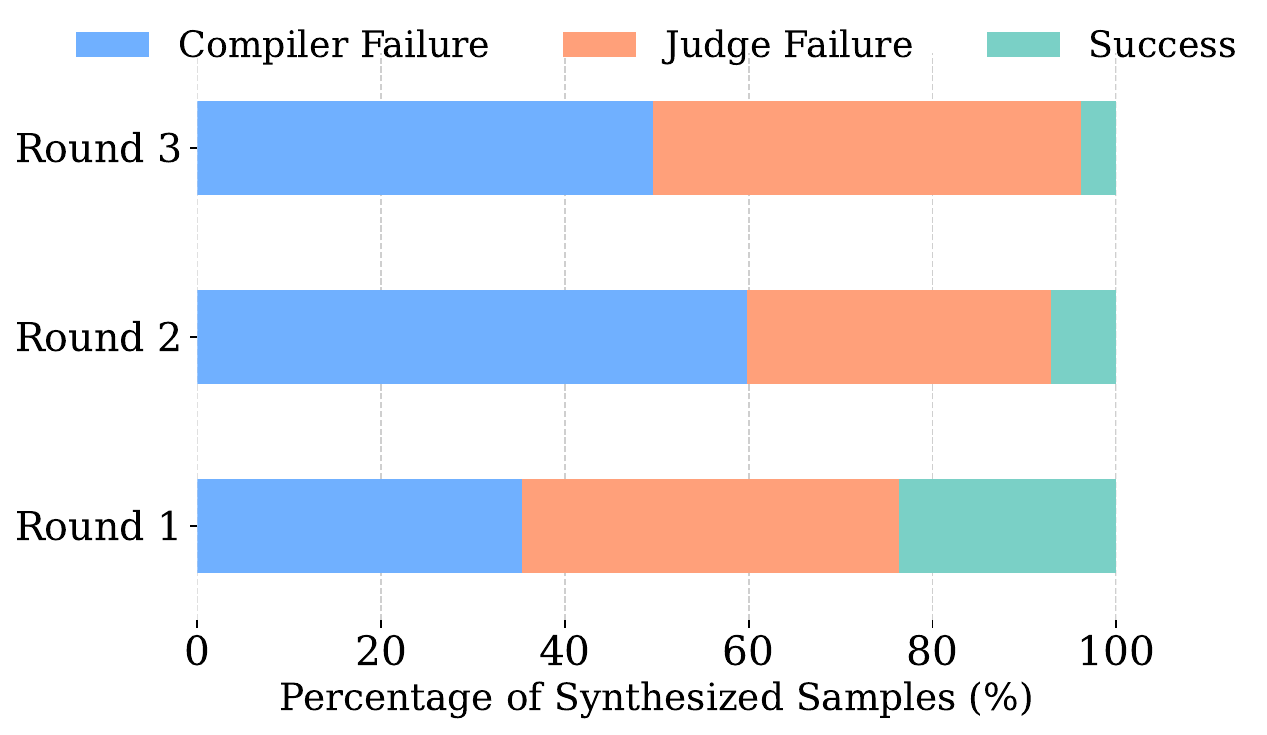}
    \vspace{-1em}
  \caption{Survival analysis of synthesized data under the medium-scale agent configuration.}
  \label{fig:survival_analysis}
      \vspace{-1em}
\end{figure}

We analyze the data survival rate,
defined as the proportion of newly synthesized samples that pass both compiler checks and LLM-based evaluation. This analysis uses the medium-scale configuration, so it characterizes the per-round yield of the framework rather than the tier composition of \ourdata, which is built with \gptoss agents.
As shown in Figure~\ref{fig:survival_analysis}, only a small fraction of the data is finally retained, and the survival rate steadily decreases across synthesis rounds.

This decline is both expected and desirable. First, the generated instructions become progressively more challenging,
reaching the limits of the agent’s capability.
Second, unlike prior work that accepts instruction-code pairs after a single pass, 
we selectively retain only high-quality, grounded data.

\section{Conclusion}
\label{sec:conc}

In this work, we introduce \ours, a dual-agent framework that orchestrates schema-driven interleaved synthesis and hybrid feedback to generate high-quality, grounded instruction-code pairs. By synergizing execution feedback with semantics,
\ours ensures both functional correctness and logical coherence.
Building on this,
we provide \ourdata,
a large-scale dataset featuring stepped difficulty levels and capturing the complete evolutionary trajectory of the synthesis process. Extensive evaluations show that \ours consistently outperforms established baselines, demonstrating that synthesis framework's design is more decisive than sheer data volume. This work offers a scalable and verifiable path toward advancing code intelligence with minimal human supervision.

\section*{Limitations}

While \ours demonstrates the potential to overcome critical challenges in acquiring instruction-code pairs data,
it is important to acknowledge certain limitations:

\paragraph{Dependence on Backbone Model.}
As a model-based approach, the quality of data synthesized by \ours is to some extent constrained by the capabilities and biases of the backbone LLMs. 
While we demonstrate that the pipeline can be effectively adapted to smaller models, stronger backbones generally yield higher-quality NL queries and code solutions, though our analysis suggests this effect also depends on the student model.

\paragraph{Throughput Constraints.}
Compared to prior prompting-based methods, \ours introduces additional computational overhead due to multi-turn agent interactions and cross-language compilation, resulting in slower data generation throughput.

\paragraph{Test Case Quality.}
During synthesis, test cases are generated autonomously by the agent, which introduces the risk of incomplete or imperfect coverage.
This is a well-known issue in software engineering, as even human-authored test suites can fail to capture all possible failure modes.
Our hybrid feedback mitigates this by filtering out flawed cases, but even with this safeguard, it remains difficult to ensure complete correctness, especially at scale.
We leave more robust test case generation as future work.

\section*{Broader Impacts}
\ours introduces a new paradigm for synthesizing high-quality code data, with the potential to improve downstream performance across a range of code generation models. Nonetheless, it is crucial to ensure that all utilized and generated code adheres to proper licensing and avoids propagating harmful coding practices.

\section*{Information About Use Of AI Assistants}
In this submission, we employed LLMs to aid and polish writing, including grammar and typo checking, as well as for identifying related works.

\section*{Acknowledgement}
This research is supported by Shanghai Artificial Intelligence Laboratory. 
We thank the reviewers of the DL4C Workshop @ NeurIPS 2025 and ACL Rolling Review for their valuable feedback, which has helped us improve this work. We are also grateful to Fangzhi Xu for his insightful comments and assistance in refining the figures.

\bibliography{references}

\appendix
\newpage
\section{Model and Training Details}
\label{app:training_details}

\subsection{Model Setting}
To validate that our method can generalize to different LLM architectures and paradigms, we experimented with five widely used base models.

\paragraph{InternLM3-8B-Instruct}
InternLM3-8B-Instruct is a typical general-purpose large language model. It follows the architecture of its predecessor models~\citep{cai2024internlm2} and is trained on 4 trillion high-quality tokens to support superior capabilities in multiple domains. It also supports long context understanding and CoT reasoning. In our experiment, we use this model to validate that our pipeline can improve the coding performance of general-purpose instruction-tuned models.

\paragraph{Qwen2.5-Coder-7B-Instruct}

Qwen2.5-Coder-7B-Instruct~\citep{qwen25coder} is an instruction-tuned language model specifically enhanced for coding tasks. It builds upon the architecture of its general-purpose base model, Qwen2.5~\citep{qwen25}, inheriting its computational efficiency and versatile vocabulary. To ensure the integrity of code understanding and generation, the model also incorporates several special tokens explicitly designed for code block generation. The model is trained on over 18 trillion tokens and incorporated extensive post-training technique, making it an ideal test bed for us to evaluate our method on coding LLMs that is transformed from general LLMs.

\paragraph{DeepSeek-Coder-6.7B-Instruct}

DeepSeek-Coder-6.7B-Instruct~\citep{guo2024deepseekcoder} is an instruction-tuned codeLLM based on the architecture of the Deepseek model~\citep{deepseekai2024deepseekllmscalingopensource}. It is trained on a corpus of 2 trillion tokens, extracted through a meticulously designed pipeline tailored for coding data. Compared to general-purpose LLMs, it employs a relatively small vocabulary specifically optimized for code-related tasks. In our experiments, we adopt this model as a representative domain-specific LLM to evaluate the effectiveness of our method on coding-oriented models.

\paragraph{StarCoder2-7B}

StarCoder2-7B~\citep{lozhkov2024starcoder2} is a representative base model from the early era of codeLLMs. It is pretrained on 3.5 trillion tokens without any additional post-training. Similar to DeepSeek-Coder, StarCoder2 employs a customized vocabulary for code-related task. We evaluate our method on this model to assess whether our trajectory data remains effective in the absence of human alignment.

\subsection{Decoding Setting}

For reproducibility, we fix all decoding hyperparameters across synthesis runs. With gpt-oss-120B, we adopt the default ``medium'' reasoning effort and set temperature to 0.6 with top-p = 1.0. For Qwen2.5-Coder-32B-Instruct and Qwen2.5-32B-Instruct, we use temperature 0.7 with top-p = 0.95. All reported evaluation numbers come from a single run per model-benchmark pair with a fixed seed. Fixed random seeds, full prompt templates, and a complete hyperparameter manifest are released alongside the code to reproduce both the synthesis pipeline and the fine-tuning runs.

\subsection{Training Setting}

For instruction-tuned models, we adopt XTuner framework~\citep{internlm2023xtuner} to streamline training. 
For StarCoder2-7B, we employ LLaMA-Factory~\citep{zheng2024llamafactory} to conduct supervised fine-tuning.
Following previous practice and our observations, we use a learning rate of $2 \times 10^{-6}$ for more stable training. For the rest of models, we used a learning rate of $5 \times 10^{-6}$.

All of our models are trained on 8 $\times$ NVIDIA A100 80GB GPUs, with a batch size of 4 per device, and a gradient accumulation of 2 steps.

\section{Evaluation Details}
\label{app:exp_details}

\paragraph{HumanEval \& MBPP.}

HumanEval~\citep{chen2021codex} and MBPP~\citep{austin2021program} are two common code completion benchmarks for evaluating the coding capability of LLMs. To further extend these two datasets, EvalPlus~\citep{liu2023evalplus} introduced HumanEval+ and MBPP+ by adding more challenging test cases and correcting inaccurate solutions. In this study, we used both the original benchmarks (HumanEval and MBPP) and their augmented versions (HumanEval+ and MBPP+) to evaluate models trained on our data as well as baseline models. We employed the official EvalPlus implementation to evaluate both benchmarks and reported 0-shot results for all variants. Following EvalPlus, MBPP and MBPP+ are evaluated on its 378-task subset rather than the full MBPP test split.
Release under MIT License.

\paragraph{BigCodeBench.}
BigCodeBench~\citep{zhuo2025bigcodebench} is a challenging benchmark for code generation, aimed at evaluating models’ ability to interpret complex instructions and invoke diverse external libraries correctly.
Under the completion setting, each task provides a function signature and docstrings, requiring the model to generate the full function implementation. 
Under the instruction setting, models are required to generate corresponding code according to a given instruction.
A unit test is also provided to verify functional correctness. Spanning a broad range of practical programming scenarios, BigCodeBench assesses models on real-world tasks that demand precise understanding of task-specific APIs and library usage. It is released under the Apache License 2.0.

\paragraph{LiveCodeBench.}

LiveCodeBench~\citep{jain2025livecodebench} is a comprehensive coding benchmark curated from mainstream competition programming platforms. It aims to provide an up-to-date, contamination-free evaluating testbed, and is continuously updated with new versions that aggregate additional problems over time. In our experiments, we use the \textit{release\_v6} version of the dataset and evaluation code (under MIT License), which comprises 1055 problems collected between May 2023 and Apr 2025.

\section{Details of Baselines}
\label{app:baseline}

\subsection{Evol Instruct}
\label{app:evol_ins}

We follow the Evol-Instruct baseline implementation used in WizardCoder~\citep{luo2024wizardcoder}.
To ensure a fair comparison with \ours, we reproduce this baseline under the same experimental setup: \qwen is used to generate instructions, and \qwenc is used to synthesize the corresponding code solutions.
We adopt the same prompt heuristics as in the original implementation, where each seed is expected to produce five instruction–code pairs.
The same set of seed data as used in \ours is employed, and the seed instructions are included in the fine-tuning process.

\subsection{OSS-Instruct}
\label{app:oss_ins}

We leverage OSS-Instruct~\citep{wei2024magicoder} from Magicoder as another strong baseline.
Specifically, we directly use the full 75K dataset released by the authors and perform fine-tuning under the same hyperparameter settings as used for \ours.

\section{Comparison}

We provide a comparative analysis with LeDex~\citep{jiang2024ledex}, Magicoder (OSS-Instruct)~\citep{wei2024magicoder}, and SelfCodeAlign~\citep{wei2024selfcodealign} with \ours,
as shown in Table~\ref{tab:approach-comparison}.

\paragraph{Interactive Synthesis Paradigm.}
 Unlike the static or one-way generation pipelines in Magicoder and SelfCodeAlign, \ours introduces a dual-agent collaborative framework. By facilitating iterative cycles between a Coder and a Reviewer, our approach ensures superior logical coherence and task complexity that exceeds simple heuristic-based synthesis.

\paragraph{Schema-Guided Grounding.}
While OSS-Instruct relies on open-source snippets as ``inspiration,'' \ours utilizes structured Schemas to guide the interleaved evolution of instructions and code. This ensures that synthesized tasks are strictly grounded and structurally sound, effectively mitigating the semantic drift common in unconstrained generation.

\paragraph{Robust Hybrid Verification.}
Moving beyond the execution-only validation found in SelfCodeAlign and LeDex, \ours employs a hybrid verification protocol. By integrating deterministic execution feedback with LLM-based semantic auditing, our framework achieves higher data fidelity by filtering out logically inconsistent false positives.

\begin{table*}[htbp]
\centering
\resizebox{\textwidth}{!}{
\small 
\renewcommand{\arraystretch}{1.3} 
\begin{tabular}{lp{3.5cm}p{3.5cm}p{3.5cm}p{3.5cm}}
\toprule
\textbf{Approach} & \textbf{Core Framework} & \textbf{Instruction Grounding} & \textbf{Verification Protocol} & \textbf{Complexity Control} \\ \midrule
\textbf{\ours} & \textbf{Dual-agent} & \textbf{Schema + Interleaving} & \textbf{Hybrid} & \textbf{Bidirectional Adaptation} \\ 
LeDex~\citep{jiang2024ledex} & Trajectory Refinement & Debugging Feedback & Execution Verification & Fixed (Error-centric) \\ 
Magicoder~\citep{wei2024magicoder} & OSS-Inspiration & Open-source Snippets & Static/Heuristic Filtering & Heuristic (Evol-Instruct) \\ 
SelfCodeAlign~\citep{wei2024selfcodealign} & Self-distillation & Concept Extraction & Sandbox Execution & Random Sampling \\ \bottomrule
\end{tabular}
}
\caption{Comparative analysis of \ours and related synthesis frameworks.}
\label{tab:approach-comparison}
\end{table*}

\section{\ourdata Details}

\begin{figure}[htb]
  \centering
  \includegraphics[width=0.75\linewidth]{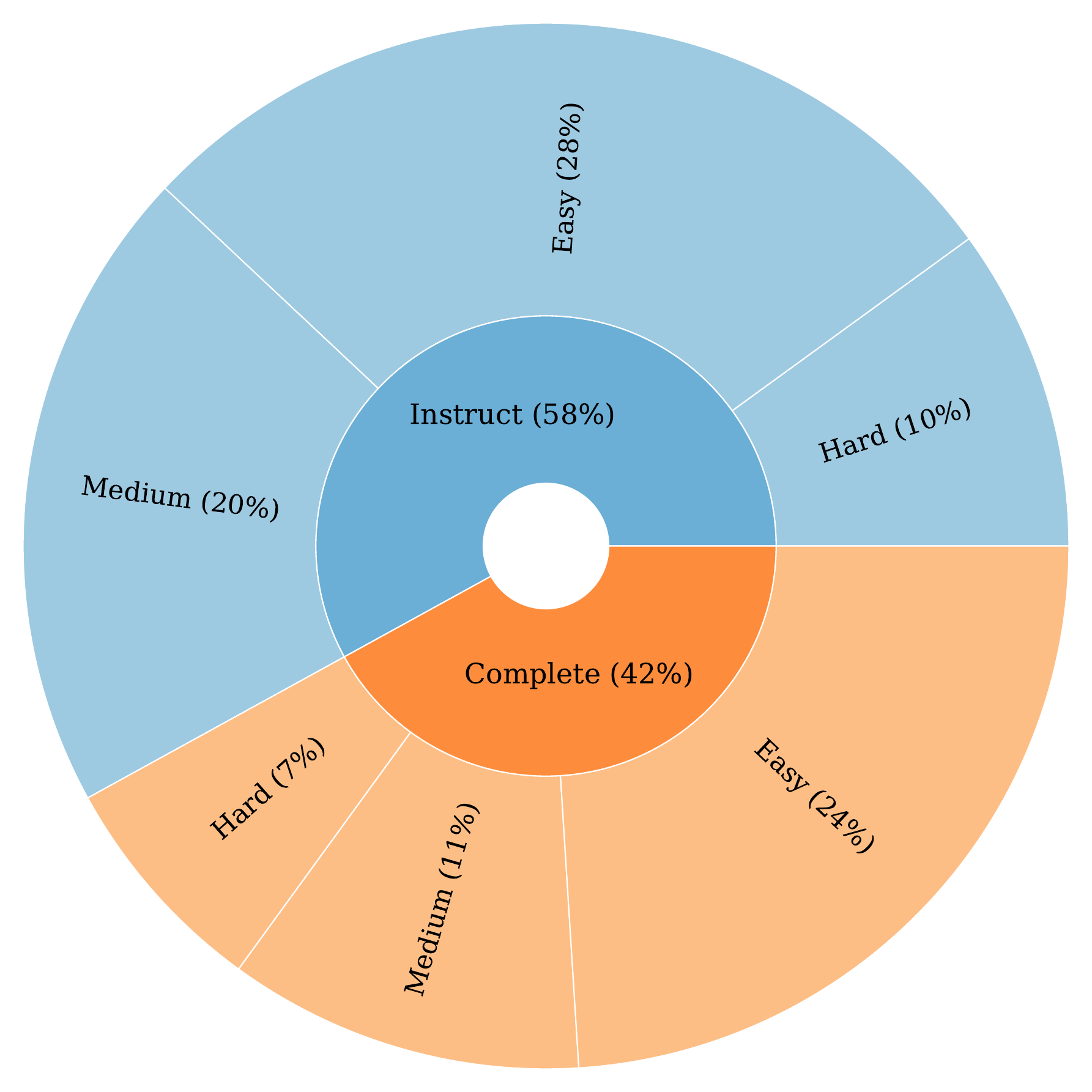}
  \caption{Composition of \ourdata by data format and difficulty tier.}
  \label{fig:dis}
\end{figure}

We detail the composition of \ourdata, as shown in Figure~\ref{fig:dis}.

It can be observed that the distributions of Instruct and Complete format data are generally comparable. Within \ourdata, 17\% of the data survived three rounds of the \ours process and forms the hard set.

\section{Code Synthesis Details}
\label{app:synthesis_details}

The prompts we used for agent collaboration are in Prompt~\ref{fig:trajectory-prompt}.

\begin{figure*}
    \centering
    \setlength{\fboxrule}{0.85pt}
    \fbox{\footnotesize
        \parbox{\dimexpr\textwidth-2\fboxsep-2\fboxrule\relax}{\textbf{Prompt for Generating CodeEvo Trajectory}\\

\textbf{Coder}: \\
Write python code to solve the following problem: \\
\{Problem description\}.\\
Include test case execution in your code.\\

\textbf{Reviewer}: \\
Next I will give you a coding problem, a piece of code, and the execution result of this code. Please determine if the code given correctly solves the problem given.\\
The problem is described as:\\
\{Problem description\}\\
The code to be assessed is:\\
\{Code from Coder\}\\
The output of this code during execution is:\\
\{Outputs from execution\}\\
The error message generated during execution is:\\
\{Errors from execution\}\\
First output "Success" or "Failure" as your judgement. Then explain the reasons and possible improvements. Do not give out improved codes.\\

\textbf{Coder}: \\
The following is an evaluation and feedback on whether the code you generated successfully answered the given question:\\
\{Feedbacks from Reviewer\}\\
Please use this feedback to improve your code so that it answers the question correctly. Still, output the refined code block only.\\

\textbf{Reviewer} \textit{(Schema planning, Eq.~\ref{eq:schema})}: \\
Below is a programming problem and a set of keywords. Design a Schema for a NEW problem that is knowledge-related to the original but more difficult. Integrate the given keywords to raise the difficulty (for example by adding constraints, less common data structures, more reasoning steps, or edge cases).\\
Output ONLY a single JSON object with exactly these fields:\\
\hspace*{1em}"concepts": a list of the key programming concepts / knowledge points the new problem combines;\\
\hspace*{1em}"combination\_logic": one sentence on how these concepts interlock in the new problem;\\
\hspace*{1em}"target\_complexity": one sentence on the intended difficulty and the reasoning or implementation steps it requires;\\
\hspace*{1em}"goal": one sentence stating what the new problem asks the solver to do.\\
Do not write the problem statement itself. Output the JSON object only.\\
Original problem:\\
\{Problem description\}\\
Keywords to integrate:\\
\{Keywords of the problem\}\\

\textbf{Reviewer} \textit{(instruction writing, Eq.~\ref{eq:write})}: \\
Given the original problem and a Schema, write the full statement of the NEW problem described by the Schema. Use the following output format:\\
\#\#\#New\\
 New programming problem you designed \\
Original problem:\\
\{Problem description\}\\
Schema (JSON):\\
\{Schema from the planning step\}\\

\textit{For the simplification branch of Algorithm~\ref{alg:codeevo-seed}, the same two prompts are used with ``more difficult'' replaced by ``SIMPLER and more tractable'' and the keywords treated as aspects to de-emphasize rather than to integrate.}\\
}
}
    \captionsetup{labelformat=default, name=Prompt}
    \caption{Prompts for generating CodeEvo Trajectory. The Reviewer first plans a Schema and then writes the new instruction from it, following Section~\ref{sec:schema}.}
    \label{fig:trajectory-prompt}
\end{figure*}

\section{Seed Instructions and Keywords}
\label{app:seed_and_keywords}

We collect instruction data from a variety of public coding platforms, including 

\begin{itemize}[itemsep=2pt,topsep=3pt,parsep=0pt]
    \item LeetCode: \url{https://leetcode.com/}
    \item Codeforces: \url{https://codeforces.com/}
    \item Codewars: \url{https://www.codewars.com/}
    \item GeeksforGeeks: \url{https://www.geeksforgeeks.org/}
    \item CodeChef: \url{https://www.codechef.com/}
\end{itemize}

We conduct a thorough similarity check and confirm that there is no contamination with the evaluation benchmarks. No personally identifiable information is present in the dataset.

The prompt used for keyword generation is provided in Prompt~\ref{fig:kw-prompt}, and the keyword sampling algorithm is detailed in Algorithm~\ref{alg:codeevo-kw}. Each seed is sampled multiple times with different keyword subsets, and every sampling is expanded under both the Instruct and Complete formats, so a single seed yields several independent evolution trajectories.

\begin{algorithm*}
\caption{Stratified Keyword Sampling Algorithm}
\label{alg:codeevo-kw}
\begin{algorithmic}[1]
\Require Label set $T$, $m = |T|$; sampling range $[r_{\min}, r_{\max}]$; maximum sampling steps $t_{\max}$
\If{$m \leq r_{\max}$}
    \For{$t = 1$ to $t_{\max}$}
        \State Randomly sample $r \sim \mathcal{U}[r_{\min}, \min(m, r_{\max})]$
        \State Sample a subset $S_t \subseteq T$, where $|S_t| = r$
    \EndFor
\Else
    \State Initialize $T_{\text{remaining}} \gets T$
    \For{$t = 1$ to $t_{\max}$}
        \If{$T_{\text{remaining}} = \emptyset$}
            \State \textbf{break}
        \EndIf
        \State Let $r \sim \mathcal{U}[r_{\min}, \min(|T_{\text{remaining}}|, r_{\max})]$
        \State Sample $S_t \subseteq T_{\text{remaining}},\ |S_t| = r$
        \State $T_{\text{remaining}} \gets T_{\text{remaining}} \setminus S_t$
    \EndFor
\EndIf
\end{algorithmic}
\end{algorithm*}

\begin{figure*}
    \centering
    \setlength{\fboxrule}{0.85pt}
    \fbox{\footnotesize
        \parbox{\dimexpr\textwidth-2\fboxsep-2\fboxrule\relax}{\texttt{\textbf{Prompt for Generating Keywords for Seed Instructions}\\
You are given a text that includes a programming problem description and explanations of its solutions. Your task is to identify and list the key programming concepts, data structures, or algorithms that are central to solving the problem. Provide your answer as a list of keywords or tags (e.g., "Array", "Hash Table", "Sorting", "Recursion", "Loop", "String", "Stack") that best capture the main ideas or techniques involved.\\
For example, if the problem involves finding two numbers in an array that add up to a target sum, appropriate tags might be "Array" and "Hash Table". \\
Now, here is the text: \\
\{text\} \\
Please provide the keywords for this problem as a comma-separated list (e.g., "Array, Hash Table"). 
        }
    }}
    \captionsetup{labelformat=default, name=Prompt}
    \caption{Prompts for generating keywords for instructions.}
    \label{fig:kw-prompt}
\end{figure*}

\section{Details of Ablation Studies}
\label{app:ablation}

Table~\ref{tab:seed_ablation_complete} reports the seed ablation for data synthesized by \gptoss agents.
Table~\ref{tab:seed_ablation_full} reports the corresponding ablation for the medium-scale setting (\qwenc as Coder, \qwen as Reviewer), covering additional backbones and benchmarks.

\begin{table*}[!h]
\centering
\resizebox{0.9\textwidth}{!}{
\begin{tabular}{lr|cc|cc|cc|cc}
\toprule
\multirow{2}{*}{\textbf{Method}} & 
\multirow{2}{*}{\textbf{Data Scale}} & 
\multicolumn{2}{c|}{\textbf{HumanEval}} & 
\multicolumn{2}{c|}{\textbf{MBPP}} &
\multicolumn{2}{c|}{\textbf{BigCodeBench-Full}} &
\multicolumn{2}{c}{\textbf{BigCodeBench-Hard}} \\
& & 
\textit{HE} & \textit{HE+} & 
\textit{MBPP} & \textit{MBPP+} & 
\textit{Instruct} & \textit{Complete} &
\textit{Instruct} & \textit{Complete} \\
\midrule
\raisebox{0.075em}{\includegraphics[scale=0.085,valign=c]{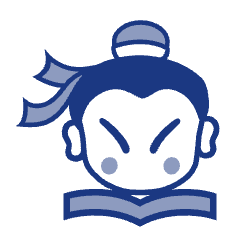}}\textit{\ivl} \\
\rowcolor{lightb} \;\ours \textit{w/o Seed} & 12K & 80.5 & 76.8 & \textbf{81.5} & \textbf{71.4} & \textbf{34.9} & 41.3 & 14.9 & \textbf{16.2} \\
\rowcolor{lightb} \;\ours & 17K & \textbf{82.3} & \textbf{78.0} & 81.2 & \textbf{71.4} & \textbf{34.9} & \textbf{43.2} & \textbf{15.5} & 15.5 \\
\seprule
\raisebox{0.075em}{\includegraphics[scale=0.12,valign=c]{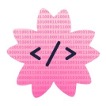}}\textit{StarCoder2-7B} \\
\rowcolor{lightb} \;\ours \textit{w/o Seed} & 12K & \textbf{51.2} & \textbf{46.3} & 64.0 & 52.9 & 29.7 & 33.8 & \textbf{8.8} & 6.8 \\
\rowcolor{lightb} \;\ours & 17K & 50.0 & 44.5 & \textbf{66.4} & \textbf{55.6} & \textbf{30.3} & \textbf{34.6} & \textbf{8.8} & \textbf{10.8} \\
\seprule
\raisebox{0.075em}{\includegraphics[scale=0.035,valign=c]{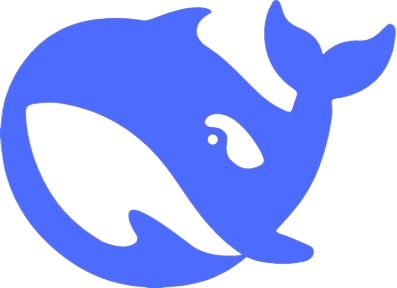}}\textit{\dpskc} \\
\rowcolor{lightb} \;\ours \textit{w/o Seed} & 12K & 76.2 & 68.9 & \textbf{77.2} & \textbf{65.9} & 36.0 & 43.4 & \textbf{12.8} & \textbf{18.6} \\
\rowcolor{lightb} \;\ours & 17K & \textbf{77.4} & \textbf{71.3} & \textbf{77.2} &\textbf{65.9} & \textbf{37.5} & \textbf{43.8} & \textbf{12.8} & 18.2 \\
\seprule
\raisebox{0.075em}{\includegraphics[scale=0.058,valign=c]{figures/icons/qwen.png}}\textit{\qwencs} \\
\rowcolor{lightb} \;\ours \textit{w/o Seed} & 12K & 84.8 & 79.3 & 78.0 & 64.8 & \textbf{42.0} & 52.1 & \textbf{17.6} & 23.6 \\
\rowcolor{lightb} \;\ours & 17K & \textbf{85.3} & \textbf{79.9} & \textbf{81.2} & \textbf{68.5} & 41.9 & \textbf{52.2} & \textbf{17.6} & \textbf{26.4} \\
\bottomrule
\end{tabular}
}
\caption{Ablation study on seed initialization. Data synthesized by the medium-scale agents (\qwenc as Coder, \qwen as Reviewer). All results are from single runs with the fixed seeds described in Appendix~\ref{app:training_details}.}
\label{tab:seed_ablation_full}
\vspace{-0.5em}
\end{table*}
\section{Extended Analysis}
\label{app:ext_analysis}

\subsection{Impact of Synthetic Data Scale}
\label{sec:analysis_scaling}

\begin{figure}[htb]
  \centering
  \includegraphics[width=1\linewidth]{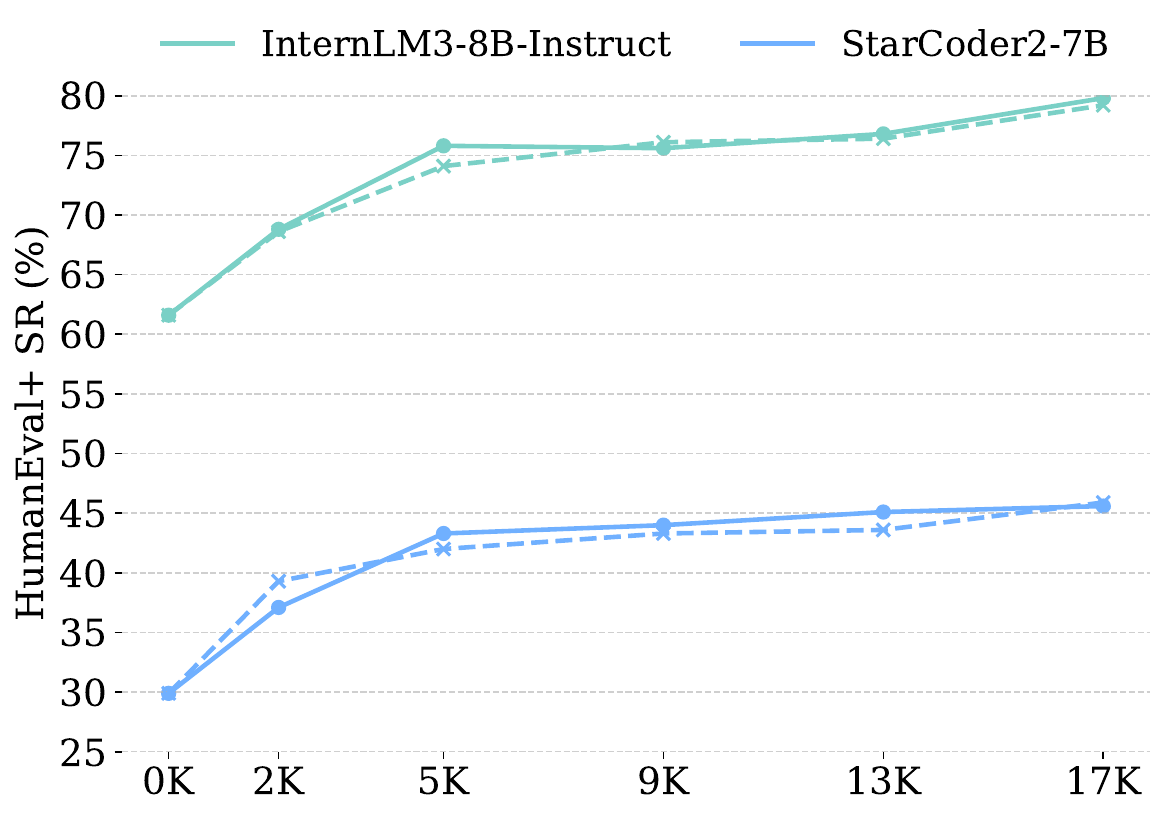}
    \vspace{-1em}
  \caption{Impact of the scale of \ours data. Solid lines and dashed lines indicate training with and without code references of seed data, respectively.}
  \label{fig:scaling}
  \vspace{-0.5em}
\end{figure}

We investigate the scaling properties of \ours-synthesized data, using 32B models as the agent backbones. 
Two models are fine-tuned with incrementally larger subsets of our synthesized dataset (along with seed data),
with the performance trend demonstrated in Figure~\ref{fig:scaling}.

It can be observed that model performance improves and then largely saturates around 5K. 
Importantly, this trend holds regardless of whether the training includes only synthetic data or a mixture of original and synthetic code. 
Within the range tested here (up to 17K), scaling \ours data introduces no distributional shift or performance degradation.

\subsection{Impact of Model Scale}
\label{app:ext_analysis:democratize}

\begin{figure}[htb]
  \centering
  \includegraphics[width=\linewidth]{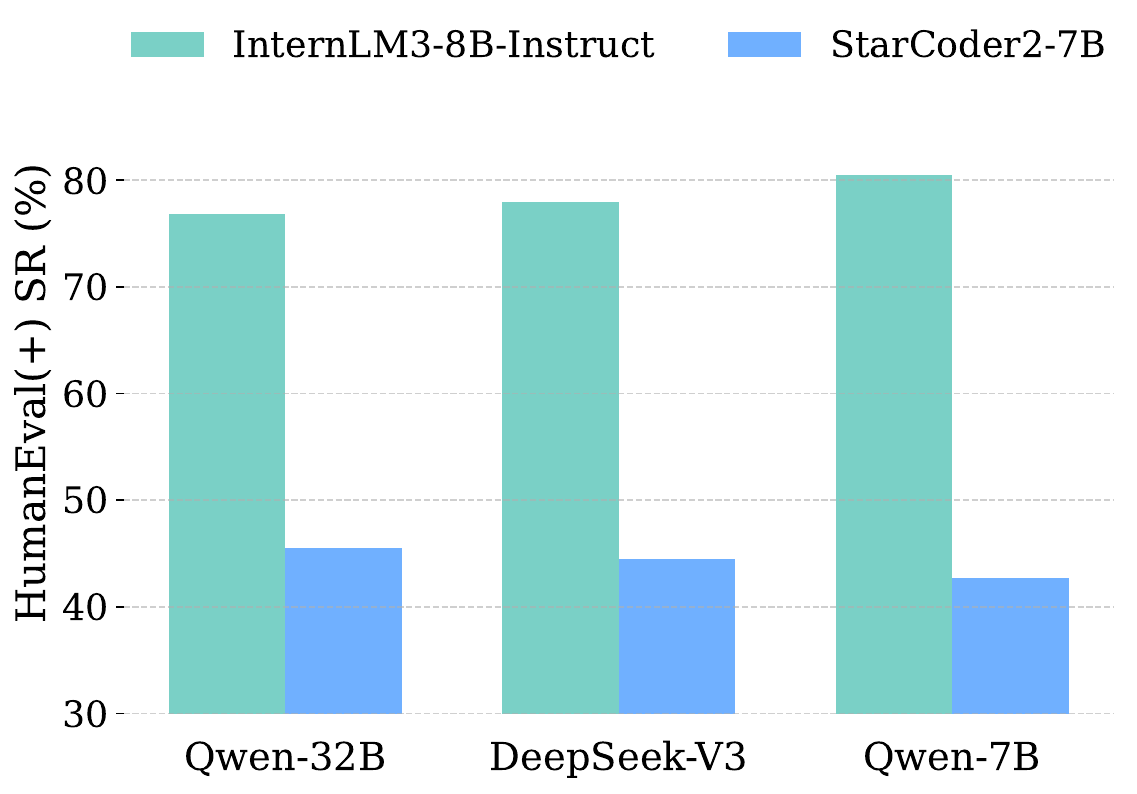}
    \vspace{-1em}
  \caption{Success rates of InternLM3 and StarCoder2 after training with data synthesized through agents with different backbones.}
  \label{fig:democratize_he}
  \vspace{-0.25em}
\end{figure}

Beyond the MBPP+ experiments, we further demonstrate the scale of agent backbones' influence on downstream results, as shown in Figure~\ref{fig:democratize_he}.

\subsection{Solvable Rate of Synthetic Instructions}
\label{app:ext_analysis:solvable}

As discussed,
a key pitfall of synthetic code data is that newly generated instructions may be ungrounded, 
\textit{i.e.}, cannot find valid solutions.
We investigate this issue by conducting a manual analysis of instructions synthesized by \ours and Evol-Instruct.

\begin{figure}[htb]
  \centering
  \includegraphics[width=\linewidth]{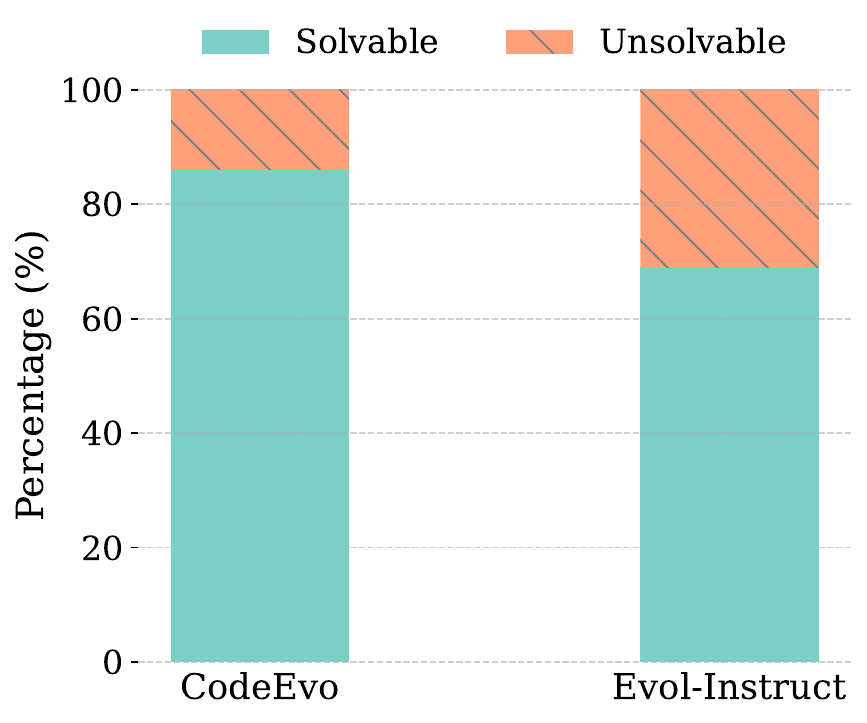}
    \vspace{-1em}
  \caption{Comparing the solvability of instructions synthesized by \ours and Evol-Instruct.}
  \label{fig:solve_rate}
  \vspace{-0.25em}
\end{figure}

The results shown in Figure~\ref{fig:solve_rate} reveal clear differences in solvability across the two approaches.
\section{Results on Additional Backbones}
\label{app:ext_results}

To further validate the effectiveness of our method, we conduct the main experiment on 3 additional backbones: \ivl~\citep{cai2024internlm2}, \starc~\citep{lozhkov2024starcoder2} and \dpskc~\citep{guo2024deepseekcoder}. We used the medium-scale agent configuration (\qwenc + \qwen) in this additional experiment. The results are shown in Table~\ref{tab:ext_python}. We can easily see that our method outperforms most of the baselines on all backbones.

\begin{table*}[!h]
\centering
\resizebox{\textwidth}{!}{
\begin{tabular}{lr|cc|cc|cc|cc|c}
\toprule
\multirow{2}{*}{\textbf{Method}} & 
\multirow{2}{*}{\textbf{Data Scale}} & 
\multicolumn{2}{c|}{\textbf{HumanEval}} & 
\multicolumn{2}{c|}{\textbf{MBPP}} &
\multicolumn{2}{c|}{\textbf{BigCodeBench-Full}} &
\multicolumn{2}{c|}{\textbf{BigCodeBench-Hard}} & 
\textbf{LiveCodeBench} \\
& & 
\textit{HE} & \textit{HE+} & 
\textit{MBPP} & \textit{MBPP+} & 
\textit{Instruct} & \textit{Complete} &
\textit{Instruct} & \textit{Complete} &
\textit{v6} \\
\midrule
\raisebox{0.075em}{\includegraphics[scale=0.085,valign=c]{figures/icons/internlm-logo.png}}\textit{\ivl} & - & 64.0 & 61.6 & 64.8 & 54.5 & 26.4 & 41.3 & 10.1 & {12.2} & 16.0 \\ 
\;Evol-Instruct & 25K & 68.3 & 64.6 & 72.4 & 62.7 & 31.5 & 39.5 & 12.8 & 14.9 & 14.9 \\
\;OSS-Instruct & 75K & 80.5 & 71.4 & 80.4 & 70.4 & 30.1 & 40.2 & 14.2 & 14.9 & 15.4 \\
\rowcolor{lightb} \;\ours & 17K & \textbf{82.3} & \textbf{78.0} & \textbf{{81.2}} & \textbf{71.4} & \textbf{34.9} & \textbf{43.2} & \textbf{15.5} & \textbf{15.5} & \textbf{17.1} \\
\seprule
\raisebox{0.075em}{\includegraphics[scale=0.12,valign=c]{figures/icons/bigcode.jpg}}\textit{StarCoder2-7B} & - & 35.4 & 29.9 & 54.4 & 45.6 & 8.8 & 10.7 & 0.6 & 4.1 & 0.6 \\
\;Evol-Instruct & 25K & 45.7 & 42.7 & 60.6 & 51.3 & 29.2 & 33.2 & 8.1 & 6.1 & 10.9 \\
\;OSS-Instruct & 75K & \textbf{50.6} & 43.9 & 60.3 & 49.7 & 29.7 & 31.4 & 7.4 & 6.8 & \textbf{12.6} \\
\rowcolor{lightb} \;\ours & 17K & {50.0} & \textbf{{44.5}} & \textbf{66.4} & \textbf{55.6} & \textbf{30.3} & \textbf{34.6} & \textbf{8.8} & \textbf{10.8} & \textbf{12.6} \\
\seprule
\raisebox{0.075em}{\includegraphics[scale=0.035,valign=c]{figures/icons/deepseek.png}}\textit{\dpskc} & - & 74.4 & 68.9 & 74.3 & 65.6 & 34.6 & 43.4 & 9.5 & 16.9 & 14.3 \\
\;Evol-Instruct & 25K & 75.0 & 68.3 & 74.9 & 64.6 & 35.8 & \textbf{44.6} & \textbf{12.8} & {16.9} & 13.1 \\
\;OSS-Instruct & 75K & 76.8 & 70.7 & \textbf{77.2} & 64.6 & 36.4 & 43.8 & 12.2 & 11.5 & 14.9 \\
\rowcolor{lightb} \;\ours & 17K & \textbf{77.4} & \textbf{71.3} & \textbf{77.2} &\textbf{65.9} & \textbf{37.5} & 43.8 & \textbf{12.8} & \textbf{18.2} & \textbf{17.7} \\
\bottomrule
\end{tabular}
}
\caption{Extended results on additional backbone models. All results are reported with pass@1(\%) performance. All results are from single runs with the fixed seeds described in Appendix~\ref{app:training_details}.}
\label{tab:ext_python}
\vspace{-0.5em}
\end{table*}

\section{Human Participants}
\label{app:human}

We recruit college-level participants with a background in computer science to conduct experiments in Section~\ref{sec:harder_ins} and Appendix~\ref{app:ext_analysis:solvable}. 
For instructions, participants are asked to follow the synthesized instructions directly as part of the evaluation process.

The instructions given to participants are shown in Prompt~\ref{fig:human-eval-prompt}. All participants are compensated at a rate of \$10 per hour for their time and effort.
We do not record any personal information, and all participants provide informed consent. The experiment does not involve surveys, interviews, or behavioral tracking.

\begin{figure*}
    \centering
    \setlength{\fboxrule}{0.85pt}
    \fbox{\footnotesize
        \parbox{\dimexpr\textwidth-2\fboxsep-2\fboxrule\relax}{\texttt{\textbf{Instructions for Human Difficulty Evaluation}\\
You are provided with a set of three programming instructions derived from the same problem: (1) some coding problems from the internet, (2) some coding problems written by LLMs, and (3) some coding problems generated by a new framework.\\
Your task is to evaluate the perceived difficulty of each instruction independently, based on your intuition as a programmer. Difficulty refers to how challenging you find the task described in the instruction, considering both problem complexity and clarity.\\
Please rate each instruction on a scale from 1 (very easy) to 5 (very difficult).\\
You do not need to write code or solve the problem—just focus on how difficult it seems to you!\\
Rating scale: \\
1 = Very easy (e.g., straightforward and simple logic)\\
2 = Easy (e.g., minor reasoning or implementation effort)\\
3 = Moderate difficulty (e.g., standard algorithm, some complexity)\\
4 = Hard (e.g., non-trivial logic or multi-step reasoning)\\
5 = Very difficult (e.g., involves advanced reasoning and multiple functions)
        }
    }}
    \captionsetup{labelformat=default, name=Prompt}
    \caption{Instructions given to human annotators for difficulty evaluation.}
    \label{fig:human-eval-prompt}
\end{figure*}

\section{Case Studies}
\label{app:cases}

We provide two case studies of \ours generating new instructions,
as shown in Prompt~\ref{fig:case-study-1} and Prompt~\ref{fig:case-study-2}.

\begin{figure*}
    \centering
    \setlength{\fboxrule}{0.85pt}
    \fbox{\footnotesize
        \parbox{\dimexpr\textwidth-2\fboxsep-2\fboxrule\relax}{\texttt{\textbf{Case of building programming problems}\\
\textbf{Seed Problem:}\\
A permutation of an array of integers is an arrangement of its members into a sequence or linear order.\\
For example, for arr = [1,2,3], the following are all the permutations of arr: [1,2,3], [1,3,2], [2, 1, 3], [2, 3, 1], [3,1,2], [3,2,1].\\
The next permutation of an array of integers is the next lexicographically greater permutation of its integer. More formally, if all the permutations of the array are sorted in one container according to their lexicographical order, then the next permutation of that array is the permutation that follows it in the sorted container. If such arrangement is not possible, the array must be rearranged as the lowest possible order (i.e., sorted in ascending order).\\
For example, the next permutation of arr = [1,2,3] is [1,3,2].\\
Similarly, the next permutation of arr = [2,3,1] is [3,1,2].\\
While the next permutation of arr = [3,2,1] is [1,2,3] because [3,2,1] does not have a lexicographical larger rearrangement.\\
Given an array of integers nums, find the next permutation of nums.\\
The replacement must be in place and use only constant extra memory.\\
\textbf{Generated Problem 1:}\\
Given an array of integers `nums`, find the next lexicographically greater permutation of its integers. If such an arrangement is not possible, rearrange the array in ascending order. Additionally, ensure that the solution handles arrays with duplicate elements efficiently and outputs the result in a circular format, where the last element points back to the first element.\\
\textbf{Generated Problem 2:}\\
Given a doubly linked list of integers `nums`, find the next lexicographically greater permutation of its integers. If such an arrangement is not possible, rearrange the list in ascending order. The solution must handle lists with duplicate elements efficiently and output the result in a circular format, where the last node points back to the first node and vice versa. Additionally, the solution should be able to handle lists with up to 1000 nodes, where each node's value is between 0 and 1000.
}
}}
\captionsetup{labelformat=default, name=Prompt}
    \caption{Case Study.}
    \label{fig:case-study-1}
\end{figure*}

\begin{figure*}
    \centering
    \setlength{\fboxrule}{0.85pt}
    \fbox{\footnotesize
        \parbox{\dimexpr\textwidth-2\fboxsep-2\fboxrule\relax}{\texttt{\textbf{Case of building programming problems}\\
\textbf{Seed Problem:}\\
You are given a string S of length 6. It is guaranteed that the first three characters of S are ABC and the last three characters are digits.\\
Determine if S is the abbreviation of a contest held and concluded on AtCoder before the start of this contest.\\
Here, a string T is "the abbreviation of a contest held and concluded on AtCoder before the start of this contest" if and only if it equals one of the following 348 strings:\\
ABC001, ABC002, \ldots, ABC314, ABC315, ABC317, ABC318, \ldots, ABC348, ABC349.\\
Note that ABC316 is not included.\\
Input\\
The input is given from Standard Input in the following format:\\
S\\
Output\\
If S is the abbreviation of a contest held and concluded on AtCoder before the start of this contest, print Yes; otherwise, print No.\\
Constraints\\
- S is a string of length 6 where the first three characters are ABC and the last three characters are digits.\\
\textbf{Generated Problem 1:}\\
You are given a string S of length 6. It is guaranteed that the first three characters of S are ABC and the last three characters are digits. Determine if S is the abbreviation of a contest held and concluded on AtCoder before the start of this contest, considering that some contest numbers were skipped due to special events. Specifically, the valid contest numbers are ABC001 through ABC314, ABC315, ABC317 through ABC348, and ABC349. Note that ABC316 is not included. Additionally, validate the input to ensure it strictly follows the format and handle any invalid inputs gracefully.\\
\textbf{Generated Problem 2:}\\
You are given a string S of length 6. It is guaranteed that the first three characters of S are ABC and the last three characters are digits. Determine if S is the abbreviation of a contest held and concluded on AtCoder before the start of this contest, considering that some contest numbers were skipped due to special events. Specifically, the valid contest numbers are ABC001 through ABC314, ABC315, ABC317 through ABC348, and ABC349. Note that ABC316 is not included. Additionally, validate the input to ensure it strictly follows the format and handle any invalid inputs gracefully. Furthermore, if the input is valid but the contest number is a multiple of 7, print "Special Case" instead of "Yes".\\
\textbf{Generated Problem 3:}\\
You are given a string S of length 6. It is guaranteed that the first three characters of S are ABC and the last three characters are digits. Determine if S is the abbreviation of a contest held and concluded on AtCoder before the start of this contest, considering that some contest numbers are missing due to special events. Specifically, the valid contest numbers are all integers from 001 to 349 except for 316. Additionally, validate the input to ensure it strictly follows the format and handle any invalid inputs gracefully by printing "Invalid Input".
}
}}
\captionsetup{labelformat=default, name=Prompt}
    \caption{Case Study.}
    \label{fig:case-study-2}
\end{figure*}

\end{document}